\documentclass[submitting]{nst}
\usepackage{amsmath}
\usepackage{subfigure,dcolumn}
\usepackage{overpic}
\usepackage[T2A,T1]{fontenc}
\usepackage[russian,english]{babel}

\newcommand{\dotx}{\dot{x}}
\newcommand{\doty}{\dot{y}}
\newcommand{\dd}{{\mathbf d}}

\newcommand{\pT} {p_{\mathrm{T}}}
\newcommand{\pTa} {p_{\mathrm{T1}}}
\newcommand{\pTb} {p_{\mathrm{T2}}}

\newcommand{\lr}[1]{\left\langle #1\right\rangle}

\newcommand{\Dphi}{\mbox{$\Delta \phi$}}

\newcommand{\beq}{\begin{equation}}
\newcommand{\eeq}{\end{equation}}
\newcommand{\ba}{\begin{array}}
\newcommand{\ea}{\end{array}}
\newcommand{\bea}{\begin{align}}
\newcommand{\eea}{\end{align}}
\newcommand{\bi}{\begin{itemize}}
\newcommand{\ei}{\end{itemize}}
\newcommand{\ben}{\begin{enumerate}}
\newcommand{\een}{\end{enumerate}}
\newcommand{\bc}{\begin{center}}
\newcommand{\ec}{\end{center}}
\newcommand{\bl}{\begin{flushleft}}
\newcommand{\el}{\end{flushleft}}
\newcommand{\br}{\begin{flushright}}
\newcommand{\er}{\end{flushright}}

\newcommand{\nn}{\nonumber \\}

\newcommand{\p}{\partial}

\renewcommand{\Re}{{\mathrm{Re}}\,}
\renewcommand{\Im}{{\mathrm{Im}}\,}

\renewcommand{\l}{\left}
\renewcommand{\r}{\right}

\usepackage{listings}
\usepackage{tikz,xcolor,hyperref}
\definecolor{lime}{HTML}{A6CE39}
\DeclareRobustCommand{\orcidicon}{
	\begin{tikzpicture}
	\draw[lime, fill=lime] (0,0) 
	circle [radius=0.16] 
	node[white] {{\fontfamily{qag}\selectfont \tiny ID}};
	\draw[white, fill=white] (-0.0625,0.095) 
	circle [radius=0.007];
	\end{tikzpicture}
	\hspace{-2mm}
}

\foreach \x in {A, ..., Z}{%
	\expandafter\xdef\csname orcid\x\endcsname{\noexpand\href{https://orcid.org/\csname orcidauthor\x\endcsname}{\noexpand\orcidicon}}
}

\begin{document}

\title{High energy nuclear physics meets Machine Learning}

\author{Wan-Bing He\orcidA{}}\thanks{Email:  hewanbing@fudan.edu.cn}
\author{Yu-Gang Ma\orcidB{}}\thanks{Email:  mayugang@fudan.edu.cn}
\affiliation{Key Laboratory of Nuclear Physics and Ion-beam Application (MOE), Institute of Modern Physics, Fudan University, Shanghai 200433, China}
\affiliation{Shanghai Research Center for Theoretical Nuclear Physics,
NSFC and Fudan University, Shanghai 200438, China}
\author{Long-Gang Pang\orcidC{}}\thanks{Email:  lgpang@mail.ccnu.edu.cn}
\affiliation{Institute of Particle Physics and Key Laboratory of Quark and Lepton Physics (MOE), Central China Normal University, Wuhan, 430079, China}
\author{Huichao Song\orcidD{}}\thanks{Email:  huichaosong@pku.edu.cn}
\affiliation{School of Physics and Center for High Energy Physics, Peking University, Beijing 100871, China}
\author{Kai Zhou\orcidE{}}\thanks{Email:  zhou@fias.uni-frankfurt.de}
\affiliation{Frankfurt Institute for Advanced Studies (FIAS), D-60438 Frankfurt am Main, Germany}

\begin{abstract}
Though being seemingly disparate and with relatively new intersection, high energy nuclear physics and machine learning have already begun to merge and yield interesting results during the last few years. It's worthy to raise the profile of utilizing this novel mindset from machine learning in high energy nuclear physics, to help more interested readers see the breadth of activities around this intersection. The aim of this mini-review is to introduce to the community the current status and report an overview of applying machine learning for high energy nuclear physics, to present from different aspects and examples how scientific questions involved in high energy nuclear physics can be tackled using machine learning. 

\end{abstract}


\maketitle

\section{Introduction}
\label{sec:intro}
Machine learning has a long history of development and application, spanning several decades. In general, it is a rapidly growing field of modern science that endows computers with the ability to learn and make predictions from data without explicit programming. It falls under the umbrella of Artificial Intelligence (AI) and is closely related to statistical inference and pattern recognition. Recently, machine learning technologies have experienced a revival and gained popularity, especially after AlphaGo from DeepMind defeated the human champion in the game of Go. This resurgence can be attributed to the advancement of algorithms, the increasing availability of powerful computational hardware such as GPUs, and the abundance of large-scale data.

Nuclear physics seeks to understand the nature of nuclear matter, including its fundamental constituents and collective behavior under different conditions, as well as the fundamental interactions that govern them. Traditional nuclear physics, especially for energies below about 1 GeV/nucleon, focuses on nuclear structures and reactions, where the degree of freedom is the nucleon. In high-energy nuclear physics, however, the degree of freedom is dominated by quarks and gluons.  Theoretical calculations and experiments or observations with large scientific infrastructures play a leading role, but are now reaching unprecedented complexity and scale. 
In the context of high energy nuclear physics, researchers are already at the forefront of \textit{big data analysis}. The detectors used in high-energy nuclear collisions, such as RHIC or the LHC, can easily produce petabytes of raw data per year. A major challenge is to make sense of the vast amounts of data generated in experiments or simulated in theory. This data is often highly complex and difficult to interpret. It's a daunting task to analyze this sheer volume of data using traditional methods of physics research. Therefore, new efficient computational methods are urgently needed to facilitate physics discovery in these computational and data-intensive research areas. 

One of the primary physical goals of high energy nuclear physics is to understand QCD matter under extreme conditions. It's expected that at extremely high temperature and/or high density, the nuclear matter, which is governed by the QCD dictated strong interaction, will turn into a deconfined quark-gluon plasma (QGP) state, with elementary particles -- quarks and gluons to be their basic degrees of freedom. The formation and properties of this new state of matter, as well as its transition to normal nuclear matter, are widely studied but still open questions in high-energy nuclear physics. This deconfined QGP state is believed to exist in the early universe, roughly a few microseconds after the big bang. Another way to study such QGP is in the course of neutron stars (or binary neutron star mergers), a compact astrophysical object whose interior serves as a cosmic laboratory for cold and dense QCD matter. Increasing astronomical observations, in particular from the progress of gravitational wave analysis, will provide more and more constraints on the extreme properties of QCD matter in this cold and dense regime, for which novel techniques to deal with the associated inverse problem will be essential.
Theoretically, first-principle lattice QCD calculations at vanishing and small baryon chemical potentials predict a smooth crossover transition from a dilute hadronic resonance gas to the deconfined QGP state. However, in the high baryon density regime, direct lattice QCD simulations are currently hampered by the fermionic sign problem. On Earth, the only chance to study this new state of QGP matter is through heavy-ion collision (HIC) programs, where two heavy nuclei are accelerated and smashed to deposit the collision energy in the overlapping region to achieve the extreme conditions, thus causing “heating/compression” on the normal nuclear matter to be excited. 

The great challenge associated with heavy ion collisions is that the collision of heavy nuclei is a highly dynamic, complex and rapidly evolving process: although the deconfined QGP state may indeed be formed during the collision, it will undergo rapid expansion and cooling for its temperature, and at some point its degrees of freedom will be reconfined into colour-neutral hadrons, which will continue to interact and decay until the detector in the experiment receives its signals. The whole collision process is too short and too small to be resolved. Experimentally, we also have no direct access to the early potentially formed QGP fireball, but only indirect measurements of the final emitted hadrons or their decay products. Furthermore, the theoretical description of the collision dynamics involves many uncertain physical factors that are not yet fully clear from theory or experimental comprehension. These uncertainties can interfere with different final physical observables in the experiment. Thus, from the limited and contaminated (i.e., heavily influenced by many uncertain factors) measurements, a reliable pinning down for the physics of the produced extreme QCD matter is non-trivial and challenging. This seriously hampers the extraction of physical knowledge from the major efforts in the heavy ion collision programs. Novel, efficient computational methods are urgently needed to address this challenge for further physics exploration.

As a modern computational paradigm, machine learning within AI has become increasingly promising in recent years for applications at the forefront of high-energy nuclear physics research. In general, machine learning algorithms can be used to automatically identify patterns and correlations in data, allowing knowledge to be extracted from data computationally and automatically. It can thus help to extract meaningful information about the underlying physics or fundamental driving laws from the available data. In contrast to the traditional focus of machine learning, which is usually predictions based on pattern recognition from the collected data, the intersection of high energy nuclear physics and machine learning is also concerned with the underlying patterns and causality for the purpose of uncertainty assessment and also physical interpretation, and thus new knowledge discovery. A recent collection of datasets from different areas of fundamental physics, including in particular high energy particle physics and nuclear physics, with supervised machine learning studies is presented in Ref.~\cite{Benato:2021olt}.

For the purpose of physics identification, the intersection of high energy nuclear physics and machine learning also goes beyond the mere application of existing learning algorithms to the data set accessible in the physics problem. Paying special attention to the physical constraints or required fundamental laws or symmetries of the systems would help the efficiency of machine learning in solving the specific physics problem. For example, when using regressive or generative models to study quantum many-body systems or general quantum field theory, implementing the symmetries of the system can greatly reduce the need for training data and improve the performance for recognition~\cite{Favoni:2020reg}. 
 It is also worth mentioning that machine learning has been applied to many topics at low and intermediate energy HICs, e.g.,~\cite{Gao2021NST,Xie2021SCPMA,Ming2022NST,Wu2022PLB,Li2021,gao2022deformation,SSPMA-2021-0299}
where a recent mini-review can be found in Ref.~\cite{He2023}, as well as hadron physics, e.g.,~\cite{had1,had2,had3}.

In addition, machine learning can also be applied in the context of simulations, which play a key role in fundamental physics research as well as in a wide range of other scientific fields such as biology, chemistry, robotics, climate modelling, etc. In high-energy nuclear physics, for both experimental and theoretical studies, simulation is an important tool, starting from the understanding of the fundamental interactions involved, e.g., in heavy ion collision dynamics and detector simulation, as well as in lattice quantum field theory simulation. Simulations are used to model the behavior of nuclear matter and its constituents, and the interactions that take place between them, which are usually highly complex with detailed use of many involved physical laws and equations or empirical phenomenological models. It has long been a major consumer of computing resources for high-statistics or high-resolution simulations of heavy ion collisions and the associated detectors in high-energy nuclear physics. The collision dynamics simulation with extensive synthetic data is required to accurately interpret the experimental measurements, which is enormously computationally and memory intensive. Machine learning can be used to improve the efficiency and descriptive power of these simulations to facilitate the physics discovery process. For example, it has been proposed to use machine learning to speed up the simulation of hydrodynamics, to optimize the parameters that go into the model simulation, to make it more robust to uncertainties, or to solve the many-body problems directly by augmenting the conventional Monte Carlo simulation.

In brief, machine learning is an effective tool that can be employed to tackle many challenges in high-energy nuclear physics. It can assist in analyzing large amounts of data from high energy nuclear physics, linking nuclear experiments to physics theory exploration effectively, optimizing simulations and calibrating models more efficiently, as well as developing new empirical and theoretical models. It is undeniable that machine learning technologies have the potential to make a significant impact, even transforming the field of high-energy nuclear physics. Therefore, it is essential to acknowledge and recognize the importance of this new paradigm in advancing the field further.

\section{Methodology}
\label{sec:method}
Machine learning (ML) can be classified in several ways. One way is to classify ML by its functionality, into classification, regression, generation and dimensionality reduction.
The other way is to classify ML by the type of training data, into supervised learning, unsupervised learning, semi-supervised learning, self-supervised learning, active learning, and reinforcement learning. For example, supervised learning requires data to be labelled in such a way that the machine can be trained to build a mapping between the input and the labels. Unsupervised learning does not need labelled data, it can learn patterns from data, assuming that the machine will make self-consistent predictions on data that is perturbed or slightly augmented. Semi-supervised learning requires a small amount of labelled data along with a large amount of unlabelled data. Self-supervised learning works with specific data such as natural language or images that are sequential. It allows the machine to predict one part of the sequence from the other part. Active learning is a type of semi-supervised learning that has two pools of data, a small pool of labelled data and a large pool of unlabelled data. The machine is trained on the labelled data and validated on the unlabelled data. The performance of the simply trained machine will be different for different samples from the unlabelled data pool. For example, the machine might be quite uncertain on one sample, predicting that the label of this sample is A with $51\%$ probability and B with $49\%$ probability. This sample is assumed to be more difficult and more important for the trained machine than simple samples where the machine's predictions are quite certain. To be more data efficient, this sample is labelled and moved from the unlabelled pool to the labelled pool for further training. Reinforcement learning uses data generated by interactions with the environment. 

Based on the previous description, the loss function for supervised learning in the regression task can be written as
\begin{align}
l = || y_{\rm pred} - y_{\rm true} ||
\end{align}
where $y_{\rm pred}=f(x, \theta)$ is the function represented by machine learning models such as decision trees or deep neural networks, $x$ is the input data and $\theta$ represents all trainable model parameters, $y_{\rm true}$ is the label of the input data $x$. The $||\cdot||$ usually represents the $l_1$ norm, which gives the mean absolute error MAE, or the $l_2$ norm, which gives the root-mean-square error RMSE.

For classification, the cross-entropy loss is widely used. It is defined as 
\begin{align}
l = - \sum_{k=1}^K p_k \log q_k 
\end{align}
where $K$ is the number of possible categories of the input data $x$, $p_k = y_{\rm true}$ is the true label (probability), $q_k = f(x, \theta)$ is the network prediction. This loss is inspired by the KL divergence, which quantifies the difference between two distributions $p$ and $q$,
\begin{align}
{\rm KL}(p || q) &= \sum_{k=1}^K p_k \log \frac{p_k}{q_k} \\
& = \sum_k p_k \log p_k - p_k \log q_k \\
& = - H(p) + H(p, q) 
\end{align}
where $H(p)$ is the entropy of the distribution $p$ and the cross entropy $H(p, q)=-\sum_k p_k \log q_k$ quantifies the average number of bits required to encode the distribution $p$ using the model $q$. 

In binary classification, the cross entropy is reduced to
\begin{align}
l = -\frac{1}{m} \sum_{i=1}^m \left[p_i \log q_i + (1 - p_i) \log ( 1 - q_i) \right].
\end{align}
where $p_i$ is the true label of the $i$'th sample whose value is 0 or 1. $q_i$ is the network prediction using the sigmoid activation function in the last layer to ensure $0 < q_i < 1$. $m$ is the number of samples in each minibatch. If the true label is $p_i = 0$, only the 2nd part contributes to the loss function.

For multi-categorical classification, the loss function is also the cross-entropy loss, with the activation function in the last layer replaced by the softmax activation,
\begin{align}
{\rm Softmax}(z_i) = \frac{e^{z_i}}{\sum_{k=1}^K e^{z_k}}
\end{align}

For unsupervised learning, the loss function can generally be written as
\begin{align}
l = ||{\rm manipulate}_1(x) - {\rm manipulate}_2(x)|| 
\end{align}
where ${\rm manipulate_{1,2}}$ represents 2 manipulations on the same data. For example, in clustering tasks, the manipulations on $x$ are to compute the total distance of samples to multiple centres. In image classification tasks, the manipulations are to compute the network prediction over two different augmentations of the same image, e.g. cropping or rotation. This loss is also called self-consistent loss.

For semi-supervised learning, the loss function is the combination of the supervised loss and the unsupervised loss,
\begin{align}
l = l_{\rm supervised} + l_{\rm unsupervised}
\end{align}

For self-supervised learning, a widely used loss function is the reconstruction loss.
For example, in computer vision, the reconstruction loss is defined as the difference
between the original image and the image reconstructed by a neural network from a masked image.
\begin{align}
l = || x - f((1 - M) \cdot x)||
\end{align}
where $x$ is the original image, $M$ is the binary mask used to remove $M=0$ pixels from an image,
$f$ is a neural network used to reconstruct the image.
The same method can be used to reconstruct natural language, 
by predicting the next sentence or missing words in a sentence.
The pre-trained network can be used in many downstream tasks such as classification, regression or generation.

In active learning, the loss function is basically the same as in supervised learning.
The difference is that the trained network ranks samples from the unsupervised pool for annotation.
So the key is to rank the samples.
There are two main ways of doing this. One way is to rank according to the entropy of the predictions made by the pre-trained network,
\begin{align}
s = - \sum_i p_i \log p_i
\end{align}
where $p_i$ is the predicted probability that the sample is in class $i$.
The other way is to rank according to the diversity of the training data set, by giving the highest rank
to the sample that has the greatest distance from the training data.

For reinforcement learning, the data is generated by subsequent interactions between the network policy and the environment. The network receives an observation $o_t$ from the environment at time $t$, makes a decision and takes an action $a_t$ on the environment, the environment returns a new observation $o_{t+1}$, an immediate reward $r_{t+1}$ and a done signal. The data are thus $\{o_t, a_t, o_{t+1}, r_{t+1}, {\rm done}\}$ trajectories.
The loss of reinforcement learning is similar to supervised learning with data $o_t$ and true labels $a_t, r_{t+1}$. 

{\it Optimisation}
The goal of machine learning is to minimize the loss of prediction on new data not used for training. In gradient-based models, this is achieved simply by stochastic gradient descent (SGD) and its variants,
\begin{align}
\theta = \theta - \epsilon \frac{1}{m}\sum_{i=1}^m \frac{\partial l_i}{\partial \theta}
\end{align}
where $\theta$ represents all the trainable parameters of the machine learning model, $\epsilon$ is a small positive number called the learning rate, $m$ is the size of the mini-batch. Updating $\theta$ with the negative gradient $-\epsilon \frac{1}{m}\sum_{i=1}^m \frac{\partial l_i}{\partial \theta}$ helps to gradually reduce the loss. This can be easily verified if there is only one trainable parameter $\theta$ and the loss is $l = \theta^2$, whose negative gradient is $-2\theta$. 

All possible values of $\theta$ form a space called the parameter space. The initial value of $\theta$ is usually a bunch of random numbers. Updating $\theta$ using SGD is analogous to walking around the parameter space looking for the minimum value of the loss function. 
The loss function can be thought of as the potential surface whose negative gradients give the direction of acceleration $\vec{a}$.
In this way, simple SGD means that the position $\theta$ in parameter space is updated using the acceleration. As a result, there are two major drawbacks to using native SGD. First, if the gradient is 0, the optimisation stops immediately. Second, the network update is much faster along the direction where the gradient is large. These two drawbacks are partly solved using the momentum mechanism \cite{goh2017why} and the adaptive learning rate \cite{Kingma2014AdamAM}. 

In reinforcement learning, the goal is to maximize the accumulated rewards in the future. The optimization is a stochastic gradient ascent. In the popular policy gradient method, 
the parameters of the policy network are updated as follows,
\begin{align}
\theta = \theta + \epsilon G_t \nabla \ln \pi(a_t | o_t, \theta)
\end{align}
where $G_t = \sum_{k=t+1}^{T} \gamma^{k-t-1} r_k$ is the return representing the accumulated rewards
in the future with a discounting factor $\gamma < 1$.

{\it Auto differentiation}
The number of trainable parameters in a deep neural network is huge. In order to learn from the data, one has to compute the negative gradients of loss with respect to each of the millions or trillions of model parameters $-\frac{\partial l}{\partial \theta}$. This is intractable 
using finite difference or analytic differentiation. Finite difference has both truncation and round-off errors that cannot be controlled. Analytical differentiation will have exploding expressions for deep neural networks that are too complex to compute efficiently. In deep learning, the negative gradient is mainly computed using auto differentiation (auto-diff), which is computationally efficient but has analytical precision. 

Auto-diff has a forward mode and a backward mode. If the deep neural network is a $R^1 \rightarrow R^n$ mapping, a forward pass gives derivatives of all output variables $y_i$ with respect to the input variable $x$. On the other hand, if the network is a $R^n \rightarrow R^1$ mapping, each forward pass returns only the derivative of the output variable $y$ on one of the input variables $x_i$. In the SGD algorithm, the backward mode is much more efficient because the mapping from $\theta$ to the loss is a $R^n \rightarrow R^1$ mapping. In the following, the forward autodiff is briefly explained.

In forward mode, auto-diff is implemented by introducing a dual number for each variable,
\begin{align}
x &\rightarrow x + \dotx \dd \\
y &\rightarrow y + \doty \dd
\end{align}
where $x$ and $y$ are two variables that require gradients, $\dot{x}$ and $\dot{y}$ are the derivatives of $x$ and $y$ with respect to some variable, as mentioned above, with $\dot{x}=1, \dot{y}=0$ will give $\partial l / \partial x$ in one pass of the forward mode, setting $\dot{x}=0, \dot{y}=1$ will give $\partial l / \partial y$ in another pass of the forward model. ${\mathbf d}$ is an infinitesimal symbol satisfying ${\mathbf d}^2 = 0$, analogous to the imaginary symbol ${\mathbf I}^2=-1$. With this definition, the traditional output $z$ of each operator will be a dual number $z+\dot{z}{\mathbf d}$ whose coefficient $\dot{z}$ is the derivative of $z$, as shown below,
\begin{align}
x + \dotx \dd + y + \doty \dd &= (x+y) + (\dotx + \doty) \dd \\
x + \dotx \dd - (y + \doty \dd) &= (x-y) + (\dotx - \doty) \dd \\
(x + \dotx \dd)*(y + \doty \dd) &= x * y + (x\doty + y\dotx) \dd \\
(x + \dotx \dd) / (y + \doty \dd) &= \frac{(x + \dotx \dd)(y - \doty \dd)}{y^2 - \doty^2 \dd^2} \\
& = \frac{x}{y} + \frac{y \dotx - x\doty}{y^2} \dd \\
\end{align}
The calculations of dual numbers can easily be extended to polynomial functions,
\begin{align}
P(x + \dotx \dd) = P(x) + P'(x) \dd
\end{align}
On the computer, more complex functions such as $\sin(x)$, $\log x$ and $e^{x}$ can be approximated by polynomial functions. In principle, auto-diff works for these functions as well. In practice, these functions can be overloaded to produce output in the form of dual numbers, e.g., $\sin(x + \dotx \dd) \rightarrow \sin x + \cos x \dot x \dd$.

Because of the universal approximation capability of DNNs and the efficient and accurate auto-diff, DNNs are widely used to represent solutions of ordinary differential equations (ODE) and partial differential equations (PDE) that require gradients. In this way, many physical problems are translated into optimisation problems. This method is commonly referred to as Physically-Informed Neural Network (PINN). Compared to traditional numerical solutions, PINN is mesh-free, works for very high dimensions, is easy to implement, especially for multi-scale and multi-physics problems.

\begin{figure}
    \centering
    \includegraphics[width=\columnwidth]{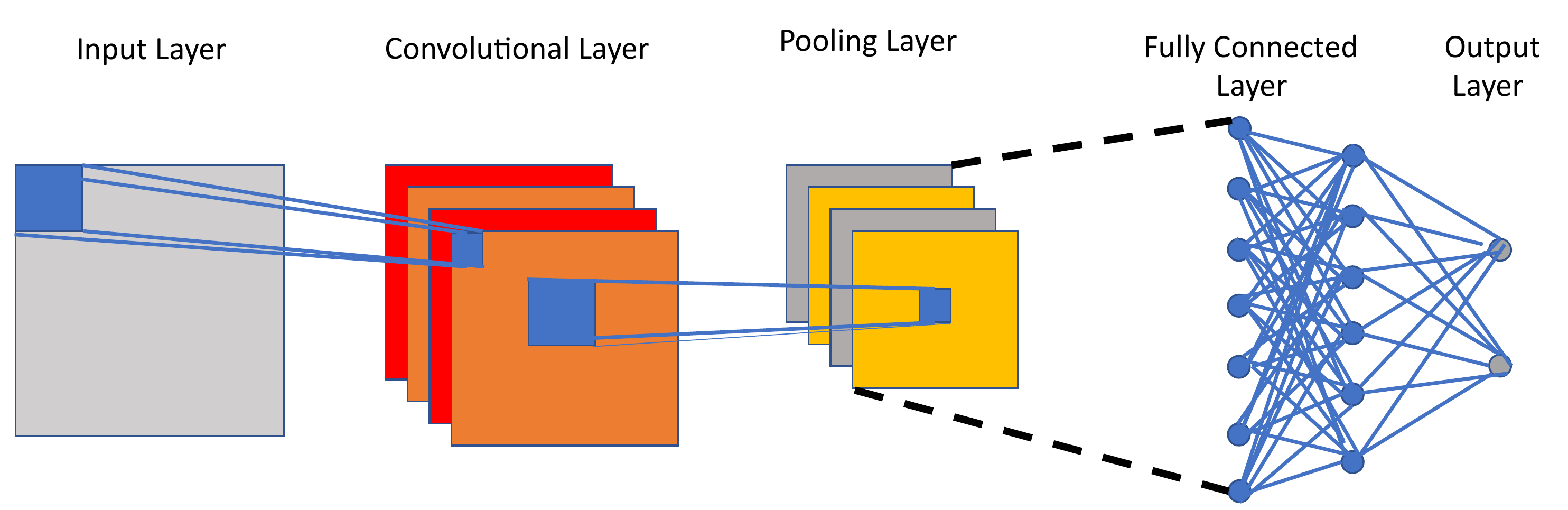}
    \caption{(Color online) Convolutional neural networks.}
    \label{fig:CNN}
\end{figure}

{\it convolutional neural networks}
Convolutional Neural Networks (CNN) are distinguished from other neural networks by their superior performance on image, speech, or audio signal inputs. A naive CNN consists of three main types of layers, i.e., convolutional layer, pooling layer, and fully-connected layer, as shown in Fig.~\ref{fig:CNN}.
The convolutional layer is the core building block of a CNN. The term convolution refers to the convolutional operation between the input features and the filters (or kernels). In the mathematical view, a convolution operation is a special kind of linear operation where two functions are multiplied to produce a third function that expresses how the shape of one function is modified by the other. In the ML view, the convolution layer uses the filters to extract the features from the input data, and combines the extracted features as the output. In a well-trained convolutional layer, a filter is only sensitive to one specific type of feature. Usually, there are many filters in a convolutional layer, to satisfy the complex input features. After the convolutional operation, a Rectified Linear Unit (ReLU) is usually chosen as the active function, which introduces nonlinearity into the neural network.

After the convolutional layer, a pooling layer is applied to reduce the number of parameters, also known as downsampling. There are two main types of pooling, that are max pooling and average pooling. Max pooling selects the maximum value to be the output, and the average pooling uses the average of the pixels covered by the pooling kernel. The fully connected layer is used to map the features extracted by the previous layers to the final output. 

The convolutional layers can be stacked to make the neural network go deeper. Earlier layers will break down the complex features from the input data to be individual simple features. As the progresses go through the next convolutional layers, the filters begin to capture larger elements or shapes of the features. With the ability to extract complex features, the CNN architecture became a foundation of modern computer vision.

However, when the neural networks go deep, the vanishing gradient problem becomes very serious. To overcome this problem in CNN architectures, many complex neural networks have been developed, such as AlexNet, VGGNet, InceptionNet, GoogLeNet, ResNet. 

{\it Recurrent neural networks}
Recurrent Neural Networks (RNN) are distinguished from other neural networks by their superior performance on sequence or time-series data.
\begin{figure}
    \centering
    \includegraphics[width=\columnwidth]{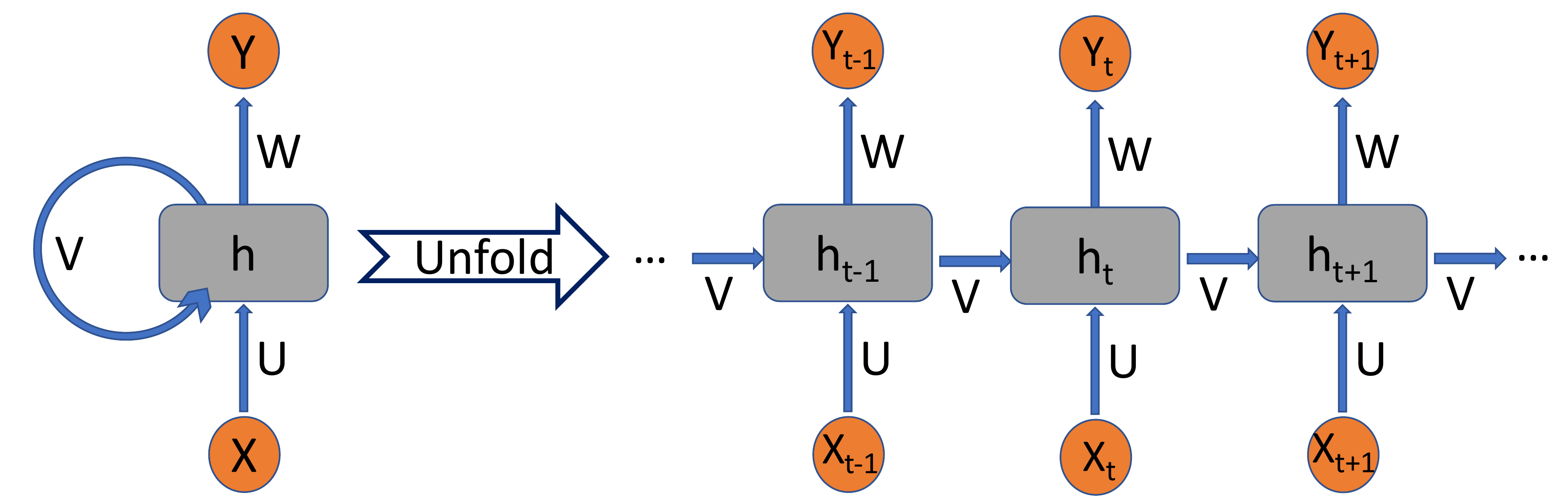}
    \caption{(Color online) Recurrent neural networks.}
    \label{fig:RNN}
\end{figure}
Fig.~\ref{fig:RNN} shows the structure of a basic RNN, where $U$ denotes the weights for the connection of the input layer to the hidden layer, $V$ denotes the weights for the connection of the hidden layer to the hidden layer, $W$ denotes the weights for the connection of the hidden layer to the output layer. Using self connection by weights $V$, RNN take information from previous inputs to influence the current input and output. This feature is often referred to as 'memory', which makes the RNN good at processing sequential data.
The loss function $\mathcal{L}$ of all time steps is defined based on the loss at each time step as follows:
\begin{equation}
    \mathcal{L}(\hat{Y},Y )=\sum_{t=1}^{T}  \mathcal{L}(\hat{Y}^{t} ,Y^{t} ).
    \label{eq:rnn_loss}
\end{equation}
The RNN uses the Backpropagation Through Time (BPTT) algorithm to determine the gradients. The error is backpropagated from the last time step to the first time step. At time step $T$, the derivative of the loss $\mathcal{L}$ with respect to the weight matrix $W$ is expressed as follows:
\begin{equation}
    \frac{\partial \mathcal{L}^{(T)} }{\partial W}=\sum_{t=1}^{T} \frac{\partial \mathcal{L}^{(t)} }{\partial W}
    \label{eq:rnn_bptt}
\end{equation}
RNNs also suffer from the problem of gradients vanishing and exploding. To deal with the gradient problems, some variant networks have been developed, such as Long Short-Term Memory Networks (LSTM) and Gated Recurrent Units (GRU).

{\it Point Cloud Network}
The final state particles from heavy ion collisions form a point cloud in momentum space.
This data must be manipulated to use CNN and RNN, as these networks were originally designed for images and natural language. 
For example, to use CNN, density estimation (histogram) is usually used to convert the particle cloud into images.
However, it does not work well for a few particles in 3-dimensional space because the particles are dilute and the resolution is poor.
To use RNN, the particle cloud must be sorted to 1 dimension, which can only keep the local information in 1d. 
The point cloud network is designed to preserve the permutation symmetry of a set of particles.
\begin{figure*}[htb]
    \includegraphics[width=15cm]{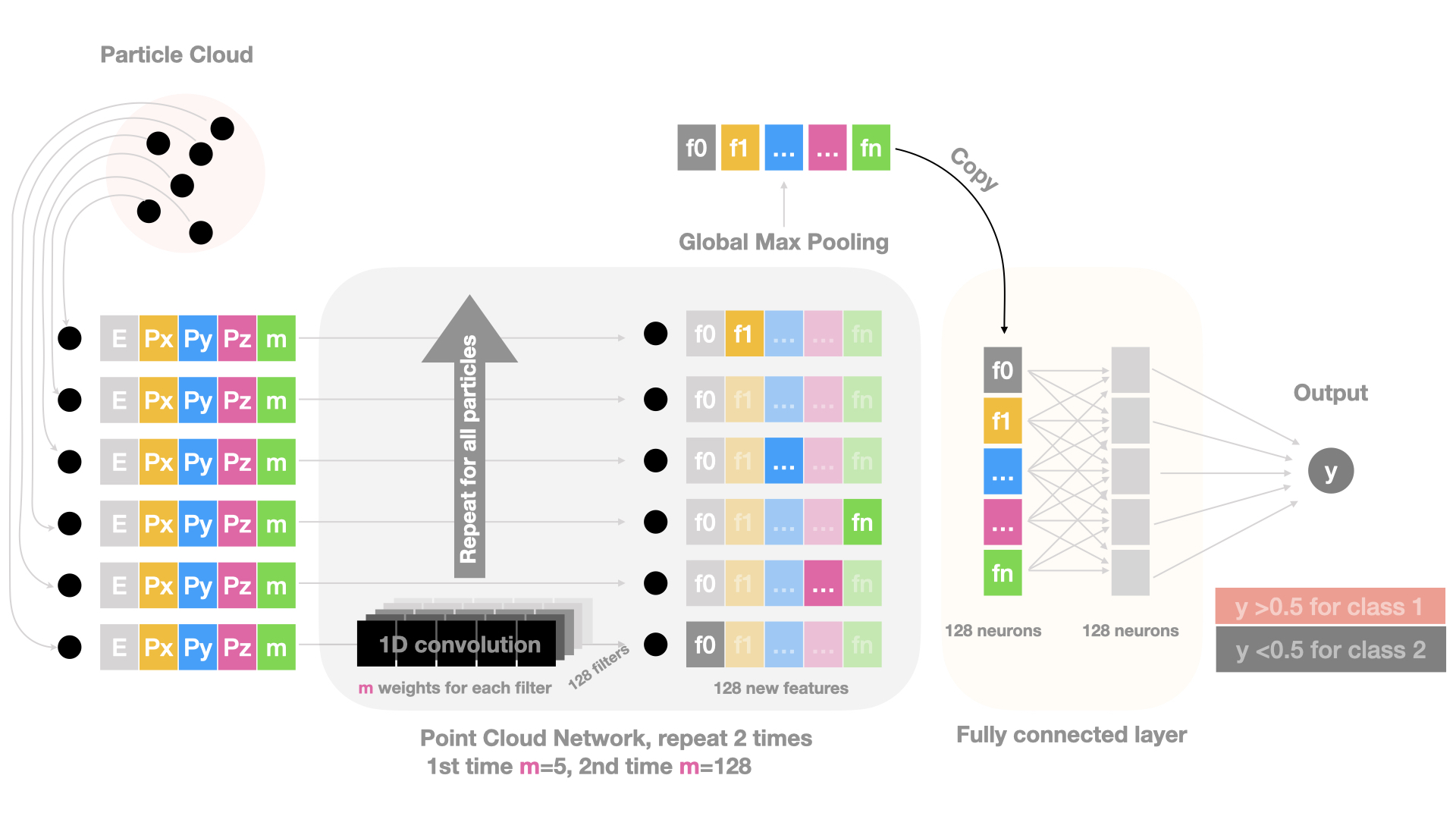}
    \caption{(Colour online) A simple demonstration of the point cloud network.}
    \label{fig:PointCloud}
\end{figure*}

Shown in figure.~\ref{fig:PointCloud} is a simple demonstration of a point cloud network.
The input to the network is a set of particles in momentum space with their 4-momentum, mass and other quantum numbers.
A fully connected neural network or Multi-Layer Perceptron (MLP) is applied to a particle to transform its m input features into 128 features in high-dimensional latent space. This MLP is shared by all particles in the cloud, which is also called a 1DCNN.
This step preserves the permutation symmetry between all particles.
Then, global max pooling (GMP) or global average pooling (GAP) is applied to these latent features of all particles,
to extract the global information of this particle cloud.
The GMP and GAP extract the boundaries of the input particle cloud in high-dimensional latent space, 
which learn the multi-particle correlation for the final decision.
This extracted global information (128 features) is further fed to another MLP for the final decision.
The output neuron has a value in the range (0, 1) and uses 0.5 as the decision boundary.

The network shown in Fig.~\ref{fig:PointCloud} is used to classify nuclear phase transitions \cite{Steinheimer:2019iso}.
Some point cloud network applies a Euclidean rotation to the point cloud to preserve rotational symmetry,
i.e., the network should make self-consistent prediction if the point cloud is rotated globally \cite{OmanaKuttan:2021axp}.
Other variants use k nearest neighbors in spatial or momentum space to extract the high dimensional latent features of each particle,
to keep more local correlation.
The k nearest neighbors of each particle can be calculated in feature space to capture long range multiple particle correlation,
because particle that are close in feature space might be far apart in spatial or momentum space.
This technique is called dynamical edge convolution and was used to look for self similarity between particles in momentum space,
which is associated to critical phenomenon that may happen in heavy ion collisions \cite{Huang:2021iux}.
The dynamical edge convolution is one kind of message passing neural network that is also called graph neural network.

{\it Generative Modelling}
In unsupervised learning, generative modelling is a class of techniques when it's related to probability distribution learning. In the sense of tasks, generally machine learning can be categorized into discriminative modelling and generative modelling. With probabilistic perspectives, discriminative modelling such as pattern recognition aim at learning a conditional probability, $p(y|x)$, which can be used to predict for a given input object ($x$) its associated properties or class identities ($y$), while the goal of generative modelling is to capture the joint distribution, $p(x,y)$, from which one can generate new data points following the same statistics as the training set. In machine learning community the generative modelling have shown great successes in numerous applications including image synthesis, inpainting, super-resolution, text-to-image translation, speech generation, chat robotics. Actually, quite a lot of the generative models were developed with profound influence from and into physics. In science there are also lots of direct applications e.g., computational fluid simulation, drug molecule design, anomaly detection, many-body physics study and also lattice field configuration generation for QCD, etc.

The central purpose of generative modelling is to get the ability to sample data ($\tilde{x}$) from the same distribution of the training set $p_d(x)$. Most of the generative modes construct parametric (explicit or implicit) models $p_{\theta}(x)$ to approach the desired data distribution.  From information theory, the Kullback-Leibler (KL) divergence (in Eq.(3)) provides an objective for this task, which measures the dissimilarity between the model and data distributions. Per Jensen inequality, the KL divergence is non-negative, and will be zero only when the two distributions match exactly. The minimization of the KL divergence under given observational data for the system with collected training set, $\mathcal{D}=\{x\}$, is equivalent to the minimization of the negative log-likelihood (NLL),
\begin{equation}
    \mathcal{L}=\frac{1}{|\mathcal{D}|}\sum_{x\in\mathcal{D}}\log p_{\theta}(x), 
    \label{eq:NLL}
\end{equation}
thus the maximum likelihood estimation (MLE).

In the following, we will review shortly several representative and popular deep generative models, including the variational autoencoder (VAE), generative adversarial networks (GAN), autoregressive modelling and normalizing flows (NF).

Variational autoencoder, VAE~\cite{Kingma2014}, 
introduces a latent variable $z$ to facilitate the generation process, thus constructs a trainable conditional probability $p_{\theta}(x|z)$ (called the decoder or generator, usually modelled by neural network). The latent variable is per generation convenience assumed to follow an easy-to-be-sampled prior distribution, $p(z)$ such as the multivariate Gaussian distribution. The introduction of latent variable however makes the data generation distribution (thus the likelihood) intractable since the needed marginalization, $p_{\theta}(x)=\int p_{\theta}(x|z)p(z)dz$. There out, the posterior distribution for the latent variable is intractable as well since $p(z|x)=p_{\theta}(x|z)p(z)/p_{\theta}(x)$. VAE employs a variational inference approach to approximately perform the maximum likelihood (MLE) on the training data. Specifically, an encoder model $q_{\phi}(z|x)$ (also modelled by a neural network) is introduced to approach the real posterior $p(z|x)$, and the KL divergence $\mathcal{D}_{KL}(q_{\phi}(z|x)||p(z|x)$ naturally provide the training objective which further derived as a variational lower bound (also known as evidence lower bound, ELBO, as the cornerstone for VAE) to the likelihood:
\begin{align}
    \mathcal{L} &= \mathbb{E}_{q_{\phi}(z|x)}[\log p_{\theta}(x|z) + \log p(z) - \log q_{\phi}(z|x)] \\
    &\le \log p_{\theta}(x), 
    \label{eq:ELBO}
\end{align}

As another latent variable generative model, the generative adversarial network, GAN~\cite{NIPS2014_5ca3e9b1}, is developed to train the generator through adversarial strategy. Intuitively, the GAN framework constructs two non-liner differentiable functions (represented both by adaptive neural network per dimensionality requesting): one called generator $G(z)$ mapping latent variable $z$ to the target data manifold $\tilde{x}=G(z)$, which gives an implicit synthesized data distribution $p_G(x)$ when the latent variable is supposed to follow a prior latent distribution $p(z)$, e.g., multivariate uniform or Gaussian, and the goal would be training the generator to push $p_G(x)$ approaching the target distribution $p_{true}(x)$; the other is called discriminator $D(x)$ which maps the data manifold to one single scalar representing the fake-vs.-true distinguish result of the discriminator on the input data. For vanilla GAN it's designed as a binary classifier for the discriminator, that for real data it's trained to output $D(x)=1$ while for generated one it's trained to output $D(\tilde{x})=0$. The generator and discriminator will be trained alternatively to improve their abilities in competing against each other, this can be achieved by mimicking a two-players min-max game, thus train the discriminator to better distinguishing the real data from the generated ones, meanwhile train the generator to cheat the discriminator for recognizing them as “real” ones.

It's proved mathematically that the adversarial training of GAN is equivalent to minimizing the JS divergence,
\begin{equation}
    \rm{JS}(p_{real}||p_G)=\frac{1}{2}(\rm{KL}(p_{real}||p_{mix})+\rm{KL}(p_G||p_{mix})), 
    \label{eq:JS}
\end{equation}
with $p_{mix}=(p_{real}+p_G)/2$. Thus, the GAN belongs to implicit MLE based generative model.The optimally trained GAN is derived to converge approaching the Nash equilibrium state, where the generator excels in synthesizing samples that the discriminator can not differentiate anymore from the real data, so the generator induced distribution indeed captured the real data distribution per training. 
This technique has been utilized in various scientific contexts, like in condensed matter physics~\cite{2017PhRvE..96d3309M, 2018PhRvE..97c2119M}, particle physics~\cite{deOliveira:2017pjk,Paganini:2017hrr}, cosmology~\cite{Ravanbakhsh:2016xpe,2019ComAC...6....1M}, also in QFT study with lattice simulation~\cite{Zhou:2018ill, Pawlowski:2018qxs}.

There are also explicit MLE based generative models, which are with close relation to statistical physics. Among them, the simplest one is autoregressive model~\cite{2015arXiv150203509G}. Basically, it invokes the probability chain rule to decompose the full probability into products of a series of conditionals,
\begin{equation}
    p_{\theta}(x)=\prod_{i}^{N}p_{\theta}(x_i|x_1,x_2,...,x_{i-1}),
    \label{eq:autoregressive}
\end{equation}
which is used as the generative model distribution to approach the desired data distribution. Specifically, one can use neural networks to parameterize each of the conditional components involved in the above. Then these neural networks as a whole actually can be viewed as one single general neural network (could take fully connected or CNN or RNN architecture) with masked weight parameter matrix (e.g., triangular with matrix for simple fully connected case) as a result of respecting the autoregressive properties specified by the Eq.~\ref{eq:autoregressive}. Being termed as PixelCNN~\cite{10.5555/3157382.3157633} or PexelRNN~\cite{10.5555/3045390.3045575} respectively, the employment of convolutional layer or recurrent layer for treating structured systems in autoregressive modelling can further respect the spatial or temporal translational invariance of the system. It also achieved state-of-the-art performances in speech synthesis with such autoregressive networks termed as WaveNet~\cite{oord2016wavenet}. With the above autoregressive representation as parametric generative model, the MLE can be explicitly performed to optimize the $p_{\theta}(x)$ for approaching the targeted data distribution $p_{real}(x)$, which as derived is minimizing the forward KL divergence $\rm{KL}(p_{real}||p_{\theta})$. This idea is also applied in many-body physics for statistical mechanics solving and general continuous system study~\cite{Wang:2020hji}.

With combination of both latent variable model and explicit maximum likelihood estimation, the normalizing flow (NF)~\cite{2014arXiv1410.8516D, 2015arXiv150505770J, 2016arXiv160508803D} is developed. Basically, NF introduces bijective affine transformations to map a simple latent space variable $z$ to the complex data manifold sample $x=g(z)$. Note that the bijectivity requires the transformation to be with the same dimensionality in input and output. This renders the usage of change of variable theorem to estimate the likelihood explicitly, 
\begin{equation}
    p_{\theta}(x)=p(z)|\det(\frac{\partial z}{\partial x})|,
    \label{eq:change-of-variable}
\end{equation}
with the determinant of Jacobian for the (inverse) transformation needed. Then, after the MLE training, the parameterized transformation just serves as a generator for new sample generation $x=g(z)$. To simplify the evaluation of needed Jacobian determinant in Eq.~\ref{eq:change-of-variable}, special network structure are adopted, e.g., those holding triangular Jacobian matrix as used in Real NVP. 
Such flow-based generative models have been implemented in lattice QFT studies~\cite{Albergo:2019eim,Kanwar:2020xzo,Boyda:2020hsi} and shown very promising indicator for further QCD study in the past few years.
Recently, such flow based model was also generalized into Fourier frequency space and use in generating Feynman paths for quantum physics~\cite{Chen:2022ytr}.

{\it Principal Component Analysis}
In machine learning, Principal Component Analysis (PCA) is a statistical technique that transforms a set of correlated variables into independent variables through orthogonal transformations.  The principal components, associated with the
obtained main eigenvectors (or non-negligible singular values)  reveal the most representative configurations of the data. As one of the unsupervised learning techniques, PCA implements the Singular Value Decomposition (SVD) on a real matrix~\cite{DBLP:journals/corr/Shlens14}:
\begin{eqnarray}
\quad\quad\quad\quad \mathbf{M}=\mathbf{{X}{\Sigma}{Z}}=\mathbf{{V}{Z}}
\end{eqnarray}
where $\mathbf{M}$ is a matrix of a size of $N\times m$, $\mathbf{{X}}$ and $\mathbf{{Z}}$ are two orthogonal matrices of the size of $N\times N$ and $m\times m$ and $\mathbf{{\Sigma}}$ is a diagonal matrix with the singular values arranged in descending order. Then, the $i_{th}$ row of the matrix $\mathbf{M}^{(i)}$ can be expressed as:
\begin{eqnarray}
\mathbf{M}^{(i)}&=&\sum_{j=1}^m {x}_j^{(i)}{\sigma}_j {z}_j
=\sum_{j=1}^m \tilde{v}_j^{(i)} {z}_j  \nonumber  \\
&\approx & \sum_{j=1}^{{k}} \tilde{v}_j^{(i)} {z}_j \ \ \ (i)=1,... ,N
\label{pca}
\end{eqnarray}
where  $\tilde{v}_j^{(i)}$ is the corresponding coefficient of ${z}_j$ for the $i_{th}$ row. In the last step, there is a cut on the indices, ${k}$ since PCA focuses only on the most important components. Due to its strong power in data mining, PCA has been widely used in various research areas of physics. For recent progress in Heavy Ion Collisions, please see \textbf{Sec.\ref{sec:obs}} in this review.

\section{Initial Condition}
\label{sec:initial}
In the traditional view, nuclear structure manifests its significance only at low energy since the high energy nucleus-nucleus collisions are a violent process in which the whole nucleus will be disassembling. However, many recent progress has demonstrated that the initial nuclear structure information is very  important for understanding the final observables in high energy heavy-ion collisions. One of the examples is collective flows, e.g., elliptic flow and triangular flow, in which the initial participant shape and nucleon density distribution as well as their initial state fluctuations are relevant. In particular, collision geometry, neutron skin, deformation, and $\alpha$-clustering structure are very important to influence the final observable. A mini-review can be found in a chapter in the handbook of nuclear physics by Ma and Zhang \cite{Ma-Zhang-Handbook}.  Machine learning provides a powerful tool for discriminating such initial structure information. In this section, we will discuss such applications. 

{\it Impact parameter estimation:}
The impact parameter $b$ describes the distance between the centers of the two colliding nuclei in the classical view, which is a crucial quantity determining the initial geometry of a collision. In experiments, the impact parameter is not directly measurable and usually estimated from the multiplicity of final-state particles in track detectors or the energy deposited in calorimeters.
Machine learning approaches are proposed to determine the impact parameters from the final-state particles, and show better performance than conventional methods. Ref.\cite{Xiang:2021ssj} proposes to use ANN and CNN to reconstruct the impact parameters from the energy spectra of final-state charged hadrons of heavy ion collisions at $\sqrt{s_{NN}}$ = 7.7 GeV to 200 GeV, which are simulated with the AMPT model. Both the ANN and CNN can reconstruct the impact parameters with a mean absolute error of about 0.4 fm. When the input feature is from a larger pseudorapidity window, the CNN shows a higher prediction accuracy than the ANN. Ref.\cite{Li:2020qqn} reports the performance of CNN and LightGBM in reconstructing the impact parameter from the heavy ion collisions at $\sqrt{s_{NN}}$ = 0.2 GeV to 1 GeV, which are simulated with the UrQMD model. The input features are constructed from the proton spectra in transverse momentum and rapidity. The average difference between the true impact parameter and the estimated one can be less than 0.1 fm. The LightGBM shows a better performance than the CNN. 

A model-independent Bayesian inference method to reconstruct the impact parameter distributions is proposed in Ref.\cite{Li:2022mni}. The impact parameter distributions are inferred from model-independent data. This method is based on Bayes' theorem,
\begin{equation}
    P\left ( b\mid \bf{X} \right ) = P\left ( b \right )P\left ( \bf{X}\mid b\right )/P\left ( \bf{X} \right ),
    \label{eq:b_bayes}
\end{equation}
$ P\left ( \bf{X} \right )$ is the probability of the observable that can be measured in the experiment. $P\left ( \bf{X}\mid b\right )$ is the probability density distribution of $\bf{X}$ for a given impact parameter b. Fluctuation is taken into account by assuming $P\left ( \bf{X}\mid b\right )$ to be a Gaussian or gamma distribution, which can be determined by fitting the data with the formula $P \left( \bf{X} \right )  = \int P \left( \bf{X} \mid b \right ) P\left(b\right )db$. $P \left( \bf{X} \right )$ can be a multidimensional form. In the Ref.\cite{Li:2022mni} two observables are used, $\bf{X}=\{M,p_{t}^{tot}\}$ , where $M$ is the multiplicity of the charged particles and $p_{t}^{tot}$ is the total transverse momentum of the light particles.

{\it End-to-end centrality estimation for CBM:}
The Compressed Baryonic Matter (CBM) detector is currently under construction for FAIR at GSI, which will study the properties of strongly compressed nuclear matter via heavy ion collisions with beam energies ranging from 2 to 10 $AGeV$. A characteristic of the CBM experiment is its very high event rate and trigger rate, which will produce a huge amount of raw data per second in real-time and pose a challenge for online event characterization and storage. To address the online event characterization, it's essential to be able to work on the direct output of the detector, which has an inherent point cloud structure--a collection of points as an unordered list with particles or tracks' attributes recorded. One important property of the point cloud is that they as a whole should be invariant under permutation. PointNet structure~\cite{8099499} is specially developed to respect this order invariance. Accordingly, to heavy ion collisions, the PointNet-based models serve as a natural way to perform real-time physics analysis on the detector output directly.

\begin{figure}
   \centering
   \includegraphics[width=0.485\textwidth]{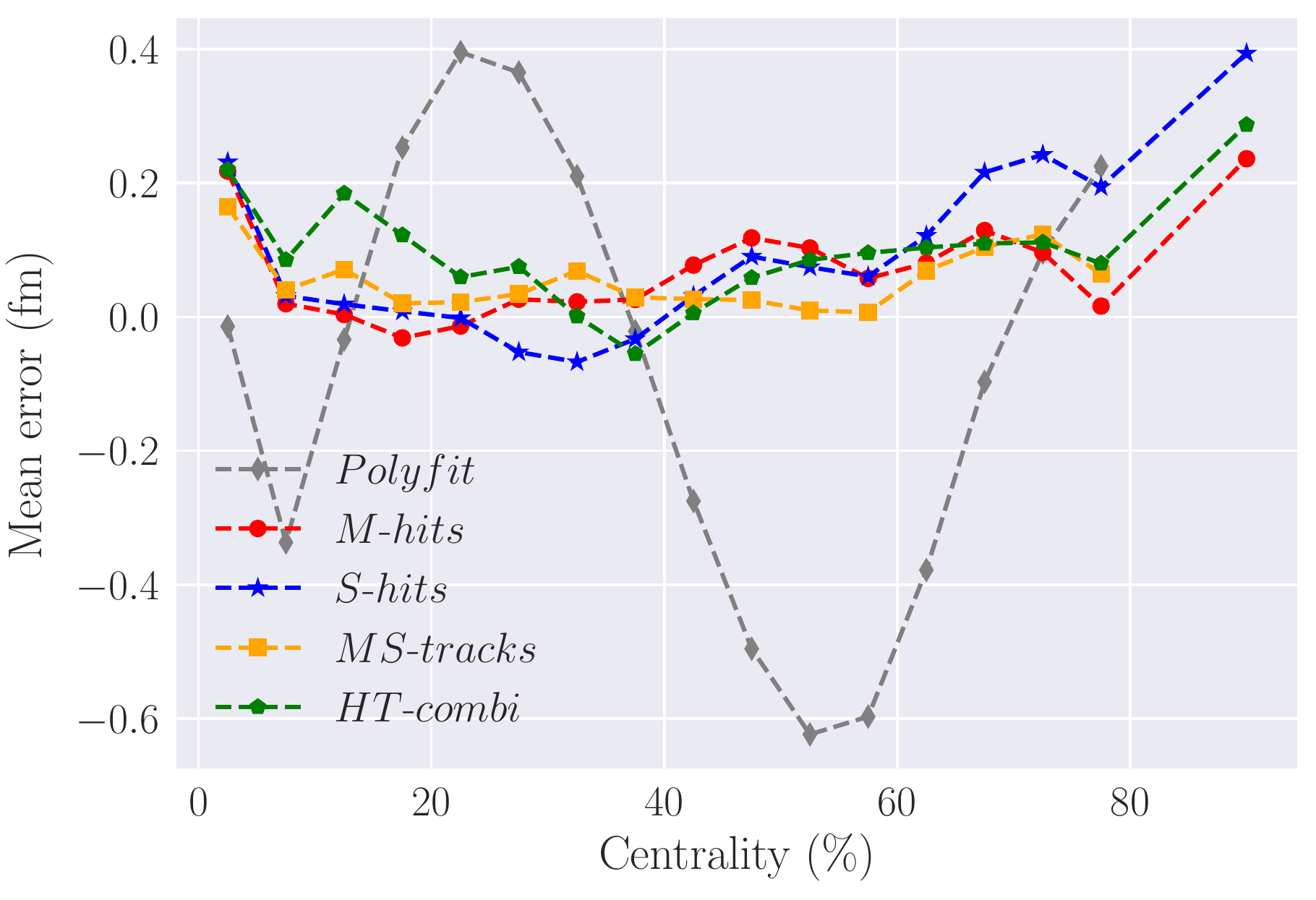}
     \caption{(Color online) Taken from Ref.~\cite{OmanaKuttan:2020brq}. Mean error in predictions as a function of centrality. Dataset \textit{Test2} is used in which peripheral events are more likely to occur. The track multiplicity is used for the centrality binning. The points at 90  $\%$ centrality are results from events with no tracks reconstructed. Therefore, the \textit{Polyfit} and \textit{MS-Tracks} model do not have a data point at 90 $\%$ centrality.}
     \label{cbm_cen}
 \end{figure}

Refs.~\cite{OmanaKuttan:2020brq, OmanaKuttan:2021axp} proposed to use PointNet based models to construct event-by-event impact parameter determination for CBM experiment using direct output from the detector, thus serving as an end-to-end centrality estimator. The supervised learning strategy is taken for this regression task, where the training data is prepared from UrQMD followed by CBMRoot detector simulation to obtain the detector output which are hits or tracks of the particles. A PointNet based model is constructed and trained to capture the inverse mapping between the detector output and the impact parameter information. It's shown that the PointNet-based model can give accurate event-by-event impact parameter determination using hits of charged particles in different detector planes and/or the tracks reconstructed from these hits, showing superior performance than a baseline model using charged track multiplicity as input inside a polynomial fit. In terms of both the precision and accuracy sector, the PointNet-based models outperform the baseline. While the baseline model gives similar resolution (relative precision) as to the PointNet-based model in the semi-central collision region, it gives a much more fluctuating and poorer accuracy for impact parameters ranging from 3 to 16$fm$, indicated by the mean of the prediction error for impact parameter. This trend also becomes more evident when it goes to realistic event distribution (i.e., $\sim bdb$), as shown by Fig.~\ref{cbm_cen} for the mean prediction error. Considering the natural parallelizibility and high speed, PointNet based model paves the way for real-time end-to-end event characterization for heavy ion collisions studies.

\begin{figure}
    \centering
    \includegraphics[width=\columnwidth]{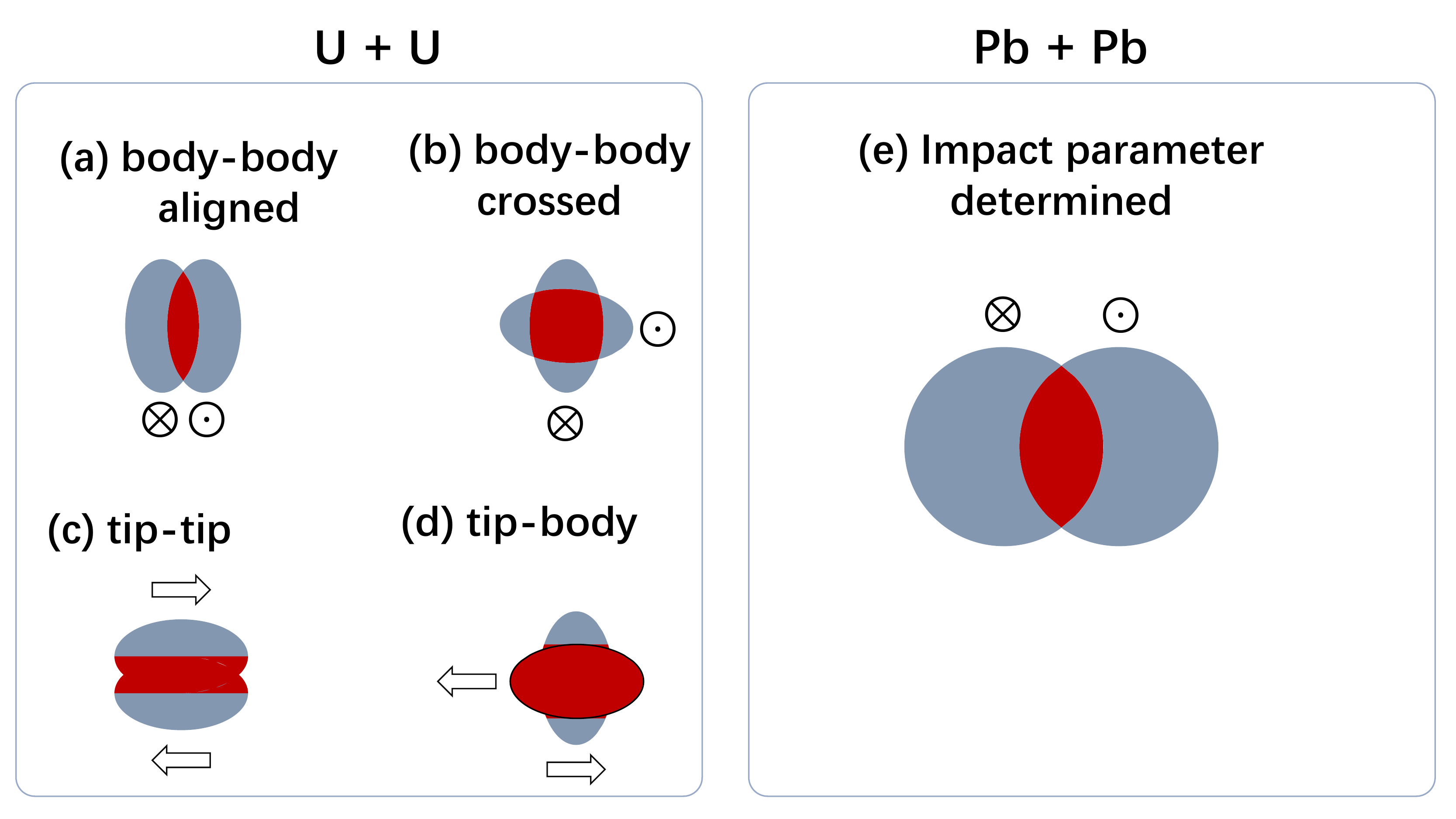}
    \caption{(Color online) Collision geometry for Pb+Pb and U+U collisions.}
    \label{fig:pbbp_vs_uu}
\end{figure}

{\it Nuclear deformation estimation:}
The momentum distribution of final state hadrons is sensitive to nuclear shape deformation.
For example, due to the different collision geometry, the elliptic flow as a function of charge multiplicity is quite different for Pb+Pb and U+U collisions.
As shown in Fig.~\ref{fig:pbbp_vs_uu}, the $^{208}{\rm Pb}$ is a double magic nucleus with an almost perfectly spherical shape, whose collision patterns depend only on the impact parameter $b$.
While the shape of $^{238}{\rm U}$ is close to a watermelon, whose collision patterns are much more complex than Pb+Pb collisions. 
For example, the U+U collisions have body-body aligned, body-body crossed, tip-tip and tip-body collisions. Different collision patterns correspond to different charge multiplicity and elliptic flow.
Both the fully overlapped body-body aligned and the central tip-tip collisions correspond to most central collisions with high charge multiplicity, but their elliptic flows are quite different.
This kind of difference will lead to a much larger variance in the elliptic flow for most central U+U collisions,
compared to high-multiplicity Pb+Pb collisions.
In principle, the complex collision patterns will lead to many differences in the elliptic flow versus charged multiplicity diagram.
Deep learning is expected to identify these differences and make a prediction of the nuclear shape deformation parameters using these patterns.

It is demonstrated that using nuclei with different deformation parameters $\beta_2$ and $\beta_4$, high-energy heavy ion collisions can be simulated using the Trento Monte Carlo model to obtain the event-by-event total initial entropy (which is proportional to the final charged multiplicity) and the corresponding geometric eccentricity (which is approximately proportional to the elliptic flow).
A deep residual neural network is trained to predict $\beta_2$ and $\beta_4$ using the 2D images of total entropy versus eccentricity \cite{Pang:2019aqb}. 
It shows that the network is successful in predicting the absolute values of $\beta_2$ and $\beta_4$,
but fails to predict their signs using the information provided.
Using the Class Activation Map (CAM) method to map the last convolutional layer onto the input image, the authors found two regions in the image that are important for decision-making.
One is the most central collision region, which is most sensitive to the variance of the elliptical flow.

Recently, Bayesian Inference with Gaussian Process emulator is used for reconstructing the nuclear structure including deformation parameters based on heavy ion collisions measurements~\cite{Cheng:2023ciy}. As a first step exploratory study, the collision observables (charged multiplicities $N_{ch}$, elliptic flow $v_2$, triangular flow $v_3$ and mean transverse momentum $\langle p_T\rangle$) are estimated with Monte Carlo Glauber model calculation (total energy $E$, elliptic eccentricity $\epsilon_2$, triangular eccentricity $\epsilon_3$ and energy density $d_\perp$), which is argued to be able to give reasonable estimator for the ratio of observables in isobaric collision systems since cancellation of dynamics' uncertainties~\cite{Zhang:2021kxj}. Under this setup, nuclear structure reconstruction based upon both the single collision system and contrast isobaric collision system measurements are discussed. For single collision systems it's found that the Woods-Saxon parameters of nuclei can be precisely inferred from final state observables estimated with ($P$,$\epsilon_2$,$\epsilon_3$,$d_\perp$). For isobar collision systems, the simultaneous inference on the two set of nuclear structures fails with only ratio of those final observables, while the further provision of single collision system's multiplicity distribution makes high precision nuclear structure reconstruction possible. Additionally, the ratio of radial flow is found to be redundant in the presence of the ratio of elliptic flow and vise versa.  

{\it $\alpha$-clustering structure:}
Clustering structure is an exotic phenomenon in nuclei, it usually takes place in light nuclei \cite{HePRL}. In nuclear collisions for light clustering nucleus against a heavy-ion, the clustering structure could make the final state particles anisotropically distributed \cite{Ma-Zhang-Handbook,Shi:2021far,ZhangPRC}. It is crucial to determine how to extract the quantitative information about the clustering from the final observables.
In the $^{12}C$ / $^{16}O$ + $^{197}$Au collisions at relativistic energies, a machine learning method is used to retrieve evidence of the cluster structures from the azimuthal angle and transverse momentum distributions of charged pions \cite{He:2021uko}. 
In this work, a Bayesian Convolutional Neural Network (BCNN) is used. 
Except for an input layer and an output layer, the hidden layers consist of four convolutional layers and three fully connected layers.  The parameters of the three fully connected layers are sampled from learned distributions that trained by Bayesian inference.
A two-dimensional histogram of azimuthal angle versus transverse momentum is designed as input. Considering the detection efficiency in experiments, charged pions with rapidity from -1 to 1 and transverse momentum from 0 to 2 GeV/c are selected. The whole data set consists of $1.6 $$\times$$ 10^{6}$ histograms with 64 $\times$ 64 bins (pixels), with different labels to indicate different configurations.
\begin{figure}
    \centering
    \includegraphics[width=\columnwidth]{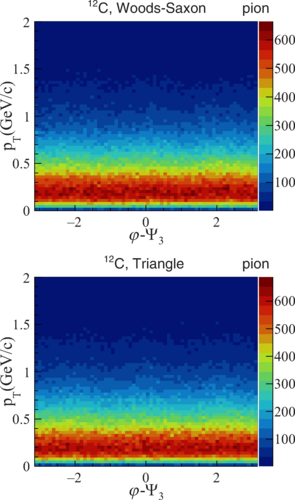}
    \caption{(Color online) Taken from Ref. \cite{He:2021uko}. Two-dimensional azimuthal angle vs transverse momentum distributions of charged pions for non-clustered (Up) and clustered (Down) $^{12}C$ from AMPT-generated $^{12}C$+$^{197}Au$ collision event at $\sqrt{S_{NN}} = 200 GeV$.}
    \label{fig:cluster-input}
\end{figure}
The typical spectra of 4000 merged events are shown in Fig. \ref{fig:cluster-input}. Even with merging, the samples of different configurations are still barely distinguishable to the naked eye. The number of merged events is denoted as NEvent, which is taken to be 1000, 2000, and 4000.
\begin{figure}
    \centering
    \includegraphics[width=\columnwidth]{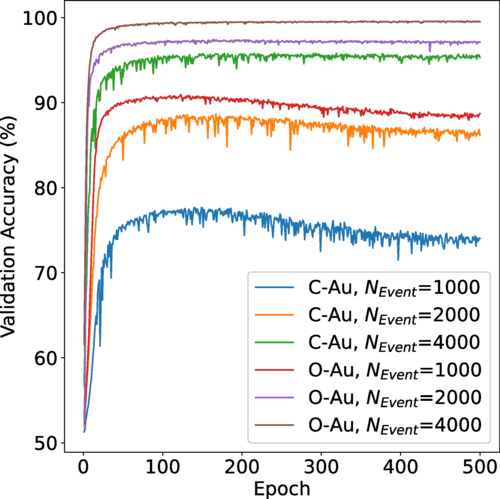}
    \caption{(Color online) Taken from Ref. \cite{He:2021uko}. Validation accuracy during the training process for colliding systems $^{12}C/^{16}O+^{197}Au$ with NEvent = 1000, 2000, and 4000.}
    \label{fig:cluster-accuracy}
\end{figure}
The learning curves are shown in Fig. \ref{fig:cluster-accuracy}. As more events are merged, the event-by-event fluctuations are reduced and the network is able to learn the features of the final state to predict the initial configuration. For $^{12}C$ with NEvent = 4000 and $^{16}O$ with NEvent = 2000, the validation accuracy reaches 95\% and 97\%, respectively, and for $^{16}O$ with NEvent = 4000 it is even 99\%.

For the cluster phenomenon, it is extremely difficult to extract signals from the final particles because fluctuations play such an important role in relativistic heavy ion collisions. By averaging over multiple events, the BCNN model is able to learn the features with promising performance.

{\it Neutron skin estimation}
The distribution of neutrons is important in determining the thickness of the neutron skin, the symmetry energy of the nucleus, the QCD EoS of dense nuclear matter, as well as 
some astrophysical observables such as the mass-radius relation of neutron stars and the gravitational wave emitted during neutron star mergers.
However, extracting the distribution of neutrons inside the nucleus is extremely difficult.
The distribution of neutrons inside the nucleus is different from protons. 
The proton distribution is much easier to measure than the neutron distribution because the former is equivalent to the charge distribution, while the latter is associated with the weak charge distribution. The neutron skin, which is the difference between the root-mean-square radius of neutrons and protons, provides a way to determine the neutron (weak charge) distribution in the nucleus.
PREX2 measured the parity-violating asymmetry by scattering longitudinally polarized electrons on Pb208 to obtain a neutron skin thickness of about $R_n-R_p=0.283\pm 0.071$ fm \cite{PREX:2021umo}.
The neutron skin is used as a constraint in the calculation of the positive and negative correlations between the symmetry energy and the slop parameter at saturation density. With this constraint, the Bayesian analysis achieves a compromise between the “conflicting” data that lead to the famous “PREXII puzzle” and the “soft Tin puzzle” \cite{Xu:2023fbs,Xu:2020fdc}. 

There are many efforts to determine the neutron skin thickness and the symmetry energy at low energy \cite{Tsang:2012se},
such as the charge-exchange spin-dipole excitation \cite{Cheng:2023yoz}, the supernova neutrinos \cite{Huang:2022wqu}, 
nuclear fragmentation reactions \cite{Teixeira:2022tnw}, 
and parity-violating electron scattering \cite{PREX:2021umo,Qweak:2021ijt}.

For high-energy heavy ion collisions, it is proposed that the isobar ratios of the charge multiplicities 
of the mean transverse momentum and the net charge multiplicities
between $^{96}_{44}{\rm Ru}+^{96}_{44}{\rm Ru}$ and $^{96}_{40}{\rm Zr}+^{96}_{40}{\rm Zr}$. 
can be used to precisely determine the nucleon skin and the symmetry energy.
The authors claim that the high-energy isobar collisions can significantly improve the result of the  
the traditional low energy method \cite{Xu:2023ges}.
In another study, the yields of spectator protons and neutrons at the forward velocity of ultra-central collisions are proposed to be good probes
of the neutron skin, sensitive to the neutron skin of $^{208}{\rm Pb}$ but insensitive to other parameters during the collision \cite{Kozyrev:2022ehy}.
An even cleaner probe is to measure the free spectator neutron yield ratios between $^{96}_{44}{\rm Ru}+^{96}_{44}{\rm Ru}$ and $^{96}_{40}{\rm Zr}+^{96}_{40}{\rm Zr}$,
in ultra-central collisions \cite{Liu:2022kvz}.

A lot of data have already been collected from high-energy heavy ion collisions.
There may be a data-driven way to reuse these data to determine neutron distribution and neutron skin thickness.
It has been tested in \cite{Huang:SSPMA-2022} that nucleons sampled from nuclei with different neutron skin types can be classified with reasonable accuracy using deep CNN and point cloud networks. However, once the nucleus is involved in heavy ion collisions, it is almost impossible to distinguish the neutron skin types of the colliding nucleus using the momentum distribution of the final state hadrons. For this particular task, the signal is weak in minimum bias collisions and deep neural networks fail to solve this difficult inverse problem. A new machine-learning method is needed to search for small and weak signals in data with large statistical fluctuations.

\section{Bulk Matter}
\label{sec:soft}

{\it{Shear and bulk viscosity}}
Shear and bulk viscosity are important properties that strongly influence the dynamical expansion of QGP and the momentum distribution of final state hadrons as shown in relativistic fluid dynamics simulations \cite{Heinz:2005bw,Romatschke:2007mq,Teaney:2003kp}. In solving the inverse problem of HIC it was found that the effects of viscosity are entangled with the initial thermalisation time, the equation of state of QGP and the phase transition between QGP and HRG. Determining the shear and bulk viscosity of hot nuclear matter thus becomes a notoriously difficult problem. Bayesian analysis plays an important role in determining the temperature dependence of the shear viscosity over the entropy density ratio $\eta/s(T)$ as well as the bulk viscosity over the entropy density ratio $\zeta/s(T)$ \cite{Bernhard:2016tnd,Bernhard:2019bmu,Yang:2022ixy}.

Suppose all parameters in the theoretical model of HIC form a set $\{\theta\}$, all experimental data from RHIC and LHC form another set $\{D\}$, then the posterior distribution of the model parameters is given by 
\begin{align}
P(\theta_i | D) =  \frac{P(D | \theta_i) P(\theta_i)}{P(D)} = \frac{P(D | \theta_i) P (\theta_i)}{\sum_j P(D | \theta_j) P(\theta_j) }
\end{align}
where $P(D | \theta_i)$ is the likelihood between experimental data $D$ and model output using parameter combinations $\theta_i$, $P(\theta_i)$ is the a priori distribution of $\theta_i$, which may be our belief based on past experience or physical considerations, the denominator $P(D)=\sum_j P(D | \theta_j) P(\theta_j)$ is a normalisation factor called evidence. Note that $P(D)$ is computationally too expensive to compute because it requires the theoretical model to traverse the entire parameter space. Fortunately, in Bayesian analysis, the normalisation factor is not really needed because the Markov Chain Monte Carlo (MCMC) method is able to sample from the following unnormalised distributions
\begin{align}
P(\theta_i | D) \propto P(D | \theta_i) P (\theta_i) 
\end{align}

The final output of the Bayesian analysis is a large number of different combinations of model parameters, sampled from the above unnormalised posterior distribution function. Doing a density estimation for each parameter, e.g., the slope of $\eta/s$ over $T_c$, gives a distribution (or histogram) of the slope parameter. This distribution has its maximum value, the location of which corresponds to the MAP estimate. It also has a variance that corresponds to the uncertainty in the slope parameter, which comes from the experimental data, the prior distribution and the likelihood function. It is thus clear that the extracted model parameters are well constrained when their posterior distribution has a narrow peak.

To estimate the temperature dependence of shear and bulk viscosity, two parameterised functions based on physical a priori are required.
In the Nature paper \cite{Bernhard:2019bmu} the shear and bulk viscosity are parameterised as,
\begin{align}
(\eta / s)(T) & =(\eta / s)_{\min }+(\eta / s)_{\mathrm{slope}}\left(T-T_{\mathrm{c}}\right)\left(\frac{T}{T_{\mathrm{c}}}\right)^{(\eta / s)_{\mathrm{crv}}} \\
(\zeta / s)(T) & =\frac{(\zeta / s)_{\max }}{1+\left(\frac{T-(\zeta / s)_{T_{\text {peak }}}}{(\zeta / s)_{\text {width }}}\right)^2}
\end{align}
where $(\eta / s)_{\min}$ and $(\zeta / s)_{\max }$ are the minimum/maximum shear and bulk viscosity values to be determined,
$T_c=154$ MeV is the QCD transition temperature representing the location of the minimum in $\eta/s(T)$,
$(\zeta / s)_{T_{\text {peak }}}$ is the location of the maximum bulk viscosity to be determined.
Other parameters to be determined are the slope $(\eta / s)_{\mathrm{slope}}$ and the curvature of the shear viscosity $(\eta / s)_{\mathrm{crv}}$,
and the width of the bulk viscosity peak $(\zeta / s)_{\text {width}}$.

Without considering other parameters, these 6 parameters form a 6-dimensional parameter space.
The above Bayes formulae are used to walk in this space, with the trajectories forming a set of parameter combinations.
This is equivalent to importance sampling using the posterior distribution of these 6 parameters.
Density estimation shows that the distribution of $(\eta / s)_{\min }$ is approximately normal,
whose mean and variance give a quantitative estimate of $(\eta / s)_{\min }=0.085_{-0.025}^{+0.026}$.
An anti-correlation is observed between $(\zeta / s)_{\max }$ and $(\zeta / s)_{\text {width}}$,
indicating that it is the integral of $(\zeta / s)(T)$ that matters, not its specific form.
The analysis also shows that the experimental data used have no constraining power on the parameters $(\eta / s)_{\mathrm{crv}}$
and $(\zeta / s)_{T_{\text {peak }}}$, as there are no obvious peaks in the posterior distributions of these 2 parameters.

{\it{Nuclear temperature}}
The difficulties in studying the finite temperature properties of nuclear matter arise mainly from the preparation of a finite temperature nuclear system and the determination of its temperature.
Heavy-ion collisions provide a possible venue for studying the finite temperature properties of nuclear matter.
During the reaction, a transient excited system is formed, which can generally be regarded as a (near) equilibrium state, since the evolution of its constituent nucleons is sufficiently short compared to the global evolution.
Its temperature can be obtained from e.g. energy spectra by moving source fitting, excited state populations, (double) isotope ratios, or quadruple momentum fluctuations, etc. For a brief review, see Ref.~\cite{EPJA_JBN}.
For a reliable thermometer of HICs, we require it to be insensitive to both the collective effects and the secondary decay of unstable nuclei after the system disintegrates, which is generally difficult to achieve.
Moreover, because of the difficulty in verifying the accuracy of the apparent temperature obtained by these thermometers, it is not a trivial task to propose different ways of determining the apparent temperature and thus provide more opportunities for cross-checking.

Using machine learning techniques, the charge multiplicity distribution can be used to determine the apparent temperature of HICs at intermediate to low energies \cite{SongYD_PLB}.
Usually, the fragment charge distributions show typical changes of the hot nuclei disassembly mechanism with temperature, i.e., from evaporation mechanism at lower temperature (e.g.,  $T^{\rm model}$= 2 MeV), to multifragmentation at intermediate temperature (e.g.,  $T^{\rm model}$= 8 MeV), till vaporization at higher temperature (e.g.,  $T^{\rm model}$= 17 MeV).

A relation between the source temperature $T^{model}$ and $M_{\rm c}(Z_{\rm cf})$ can then be established by a DNN that transforms the complex relation into a nonlinear map through its neurons.
This relation can be used to determine the apparent temperature of a particular transient state during HICs at intermediate-to-low energies, such as the $^{103}{Pd}$ + $^9{Be}$ fragmentation reaction, via their final-state $M_{\rm c}(Z_{\rm cf})$.
With the final state charge multiplicity distribution simulated with the IQMD, its apparent temperature was obtained by the trained DNN.

\begin{figure}[htb]
\includegraphics[width=0.9\columnwidth]{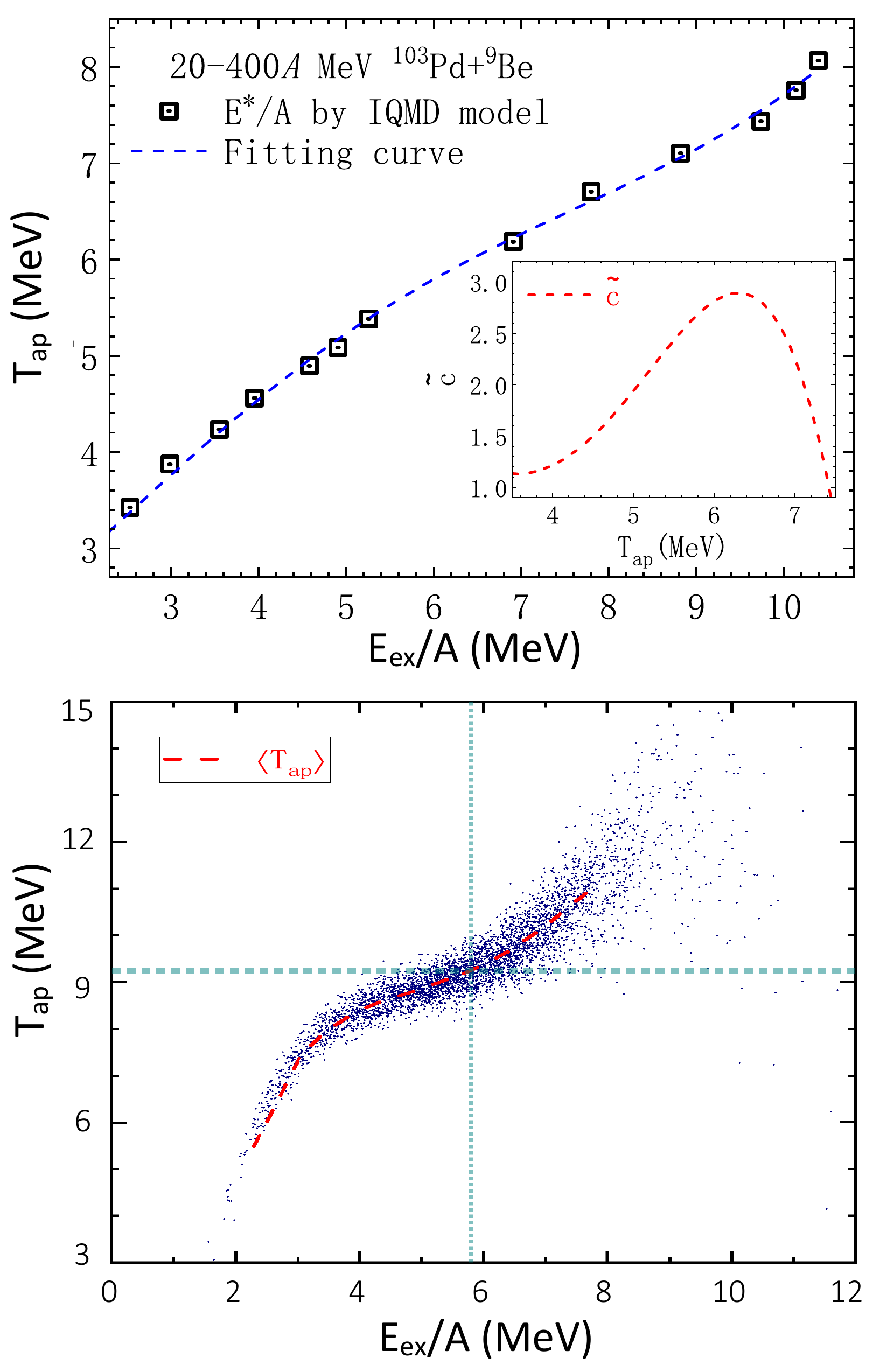}
\caption{(Color online) Upper panel: Taken from Ref. \cite{SongYD_PLB}. Caloric curve of the reaction $^{103}{Pd}$ $+$ $^9{Be}$.
Black open squares represent the result based on the IQMD model, with $T_{\rm ap}$ determined by DNN using $M_{\rm c}(Z_{\rm cf})$.
The blue dashed line is its polynomial fit.
The inset shows the specific heat capacity $\tilde{c}$ derived from the fitted formula.
Lower panel: Taken from Ref. \cite{WangR_PRR}.
Scatter plot of the apparent temperature versus the excitation energy per nucleon.
The red dashed line represents $\langle T_{\rm ap}\rangle$ as a function of $E_{\rm ex}/A$.
The horizontal and vertical cyan dotted lines represent the limit temperature and an analogical characteristic value of $E_{\rm ex}/A$ obtained by the confusion scheme, respectively (see below).
}\label{fig_song}
\end{figure}

With the $T_{\rm ap}$ determined by DNN, the caloric curve, i.e., the apparent temperature as a function of the excitation energy per nucleon $E^*/A$ of the HICs, was also examined.
Fig.~\ref{fig_song}(a) shows the caloric curve of the $^{103}{Pd}$ $+$ $^9{Be}$ reaction from the IQMD simulation with the apparent temperature determined by DNN using $M_{\rm c}(Z_{\rm cf})$ of the reaction.
As shown in the figure, the increase of $T_{\rm ap}$ slows down when $E^*/A$ reaches to about $8~\rm MeV$.
Traditionally, this characteristic behavior of the caloric curve is explained by the fact that, as the excitation energy increases, the system is driven into a spinodal region, in which part of the excitation energy begins to transfer to latent heat.
The $T_{\rm ap}$ at the maximum of the specific heat capacity of the system $\tilde{c}$ $\equiv$ $\frac{d(E^*/A)}{dT_{\rm ap}}$ is called the limit temperature $T_{\rm lim}$.
The caloric curve in Fig.~\ref{fig_song}(a) leads to $T_{\rm lim}$ $=$  $6.4~\rm MeV$ (through a polynomial fit of the $\tilde{c}$, red dashed line in the inset of Fig.~\ref{fig_song}(a)).
This value of $T_{\rm lim}$ follows the general trend of the Natowitz's limiting-temperature dependence on the system size~\cite{NatPRC65}, and thus indicates the validity of determining the apparent temperature through charge multiplicity distribution presented in this article.

\begin{figure*}[htb]
\centering
\includegraphics[width=16cm]{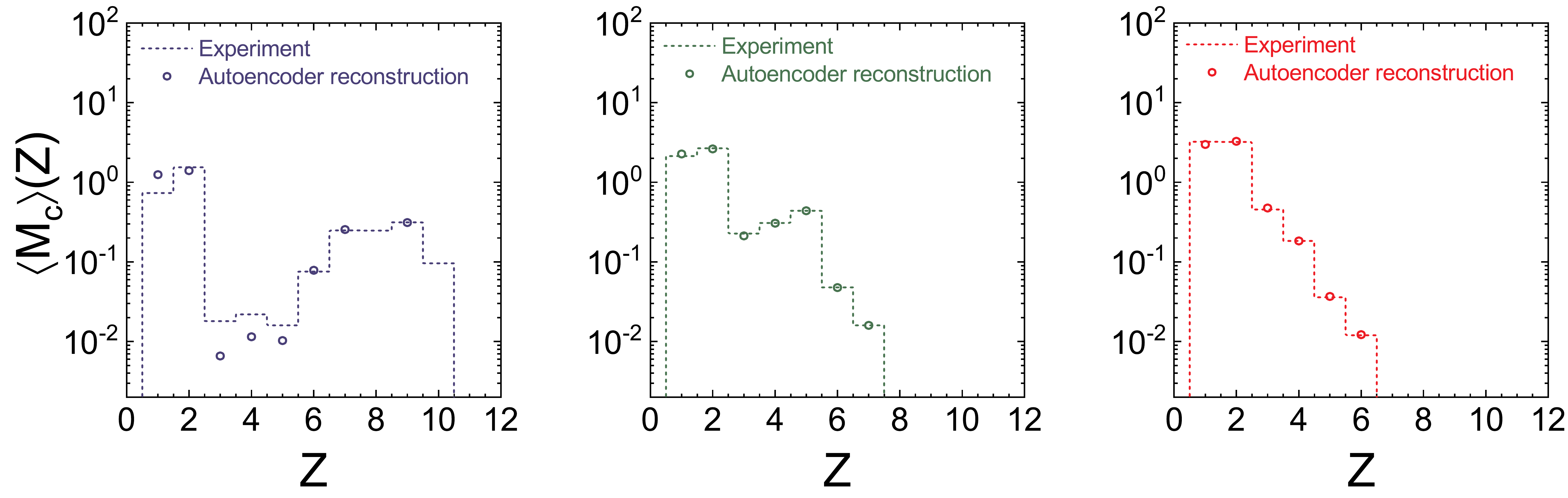}
\put(-380,83){\includegraphics[width=1.5cm]{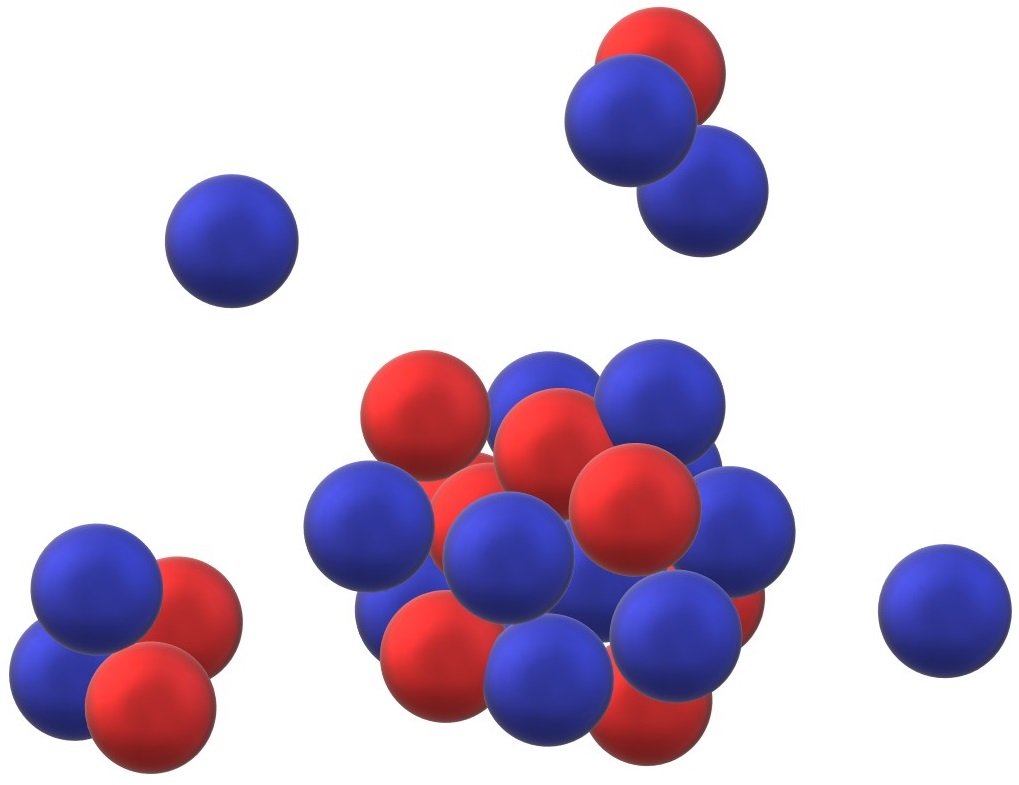}}
\put(-210,70){\includegraphics[width=1.5cm]{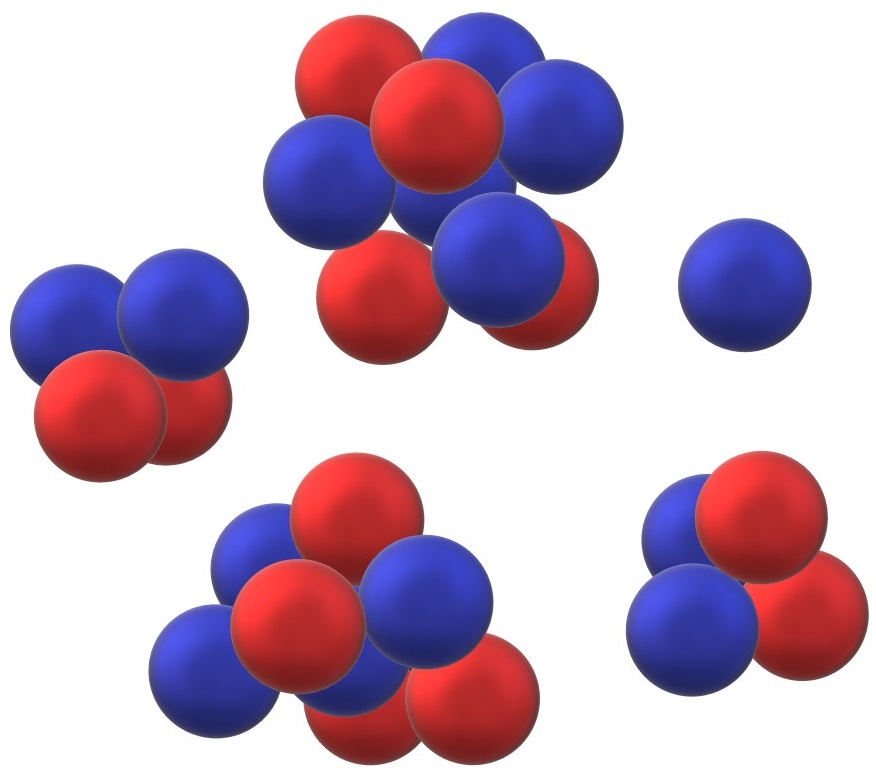}}
\put(-60,65){\includegraphics[width=1.7cm]{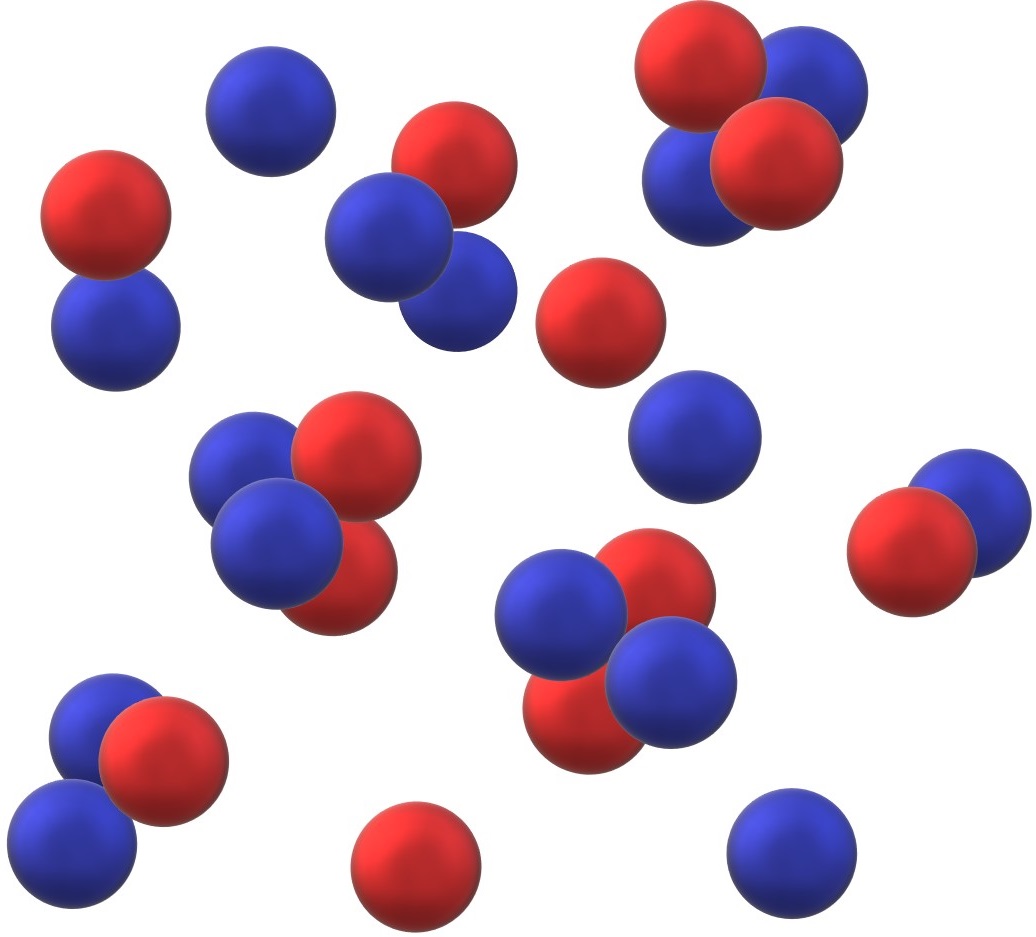}}
\put(-452,136){\bfseries (a)}
\put(-299,136){\bfseries (b)}
\put(-146,136){\bfseries (c)}
\caption{\small (Color online) Taken from Ref. \cite{WangR_PRR}. The averaged charge multiplicity distribution $\langle M_{\rm c} \rangle(Z)$ of the QP fragments.
The average is taken for different $E_{\rm ex}/A$ bins, left panel for low excitation~($0.9~\rm MeV$ - $2.8~\rm MeV$), middle panel for intermediate excitation~($5.3~\rm MeV$ - $5.4~\rm MeV$), and right panel for high excitation~($8.1~\rm MeV$ - $13.0~\rm MeV$).
The dashed curves represent $\langle M_{\rm c}\rangle(Z)$ from the NIMROD experiment, while the circles from the autoencoder network reconstruction $\langle M_{\rm c}'\rangle(Z)$.
Each $E_{\rm ex}/A$ bin contains $500$ test events.}
\label{F:CM}
\end{figure*}

{\it{Nuclear liquid gas phase transition}}
As mentioned above,  machine-learning techniques can  be used to study nuclear temperature and then the caloric curve, which is of particular interest in reaction dynamics. 
A lot of probes by analyzing sophisticatedly the information of the reaction products have been proposed to recognize the liquid-gas phase transition of nuclei~\cite{PPNP,PocPRL75,MYGPLB390,MYGPRL83,NatPRL89,NatPRC65,MYGPRC71,Deng,Liu}.
Since the nucleus is an uncontrollable system, its liquid-gas phase transition is realized by tracing the effect of the spinodal instability on the reaction dynamics, e.g., by measuring the properties of the intermediate mass fragments~(with charge number greater than $3$).
In a recent work, the averaged charge multiplicity distribution $\langle M_{\rm c} \rangle(Z)$ of the quasi-projectile (QP) fragments in the reactions of $^{40}{Ar}$ on $^{27}{Al}$ and $^{48}{Ti}$ at $47~\rm MeV/nucleon$ has been used to study the nuclear liquid-gas phase transition by the autoencoder method and a confusion scheme \cite{WangR_PRR}.
The QP fragments are supposed to come from the excited projectile nucleus, which can largely avoid the effect of dynamical evolution.
They can be obtained by a three-source (i.e., a QP source, an intermediate velocity source and a quasi-target source) reconstruction method~\cite{MYGPRC71}.
Fig.~\ref{fig_song}(b) shows the scatter plot of $T_{\rm ap}$ versus the system's excitation energy $E_{\rm ex}/A$, i.e., the caloric curve, of the events with $Z^{\rm QP} = 12$ from the experimental data.

The event-by-event charge-weighted charge multiplicity distribution of QP fragments $ZM_{\rm c}(Z)$ from the experiment are used as the input to train the autoencoder network.
The network consists of two main parts, the encoder part encodes the input event-by-event $ZM_{\rm c}(Z)$ into a \emph{latent variable}, and the decoder part decodes the latent variable into $ZM'_{\rm c}(Z)$.
The neural network is trained to recover the encoded information as best as possible, i.e., the network is trained to minimize the difference between $ZM_{\rm c}(Z)$ and $ZM'_{\rm c}(Z)$.
Fig.~\ref{F:CM} shows the averaged $M_{\rm c}(Z)$ in three typical $E_{\rm ex}/A$ bins (dashed lines).
For the test QP events, the reconstructed $M_{\rm c}'(Z)$ are averaged and compared with the original $M_{\rm c}(Z)$ in Fig.~\ref{F:CM}.
Once the autoencoder network is trained, each QP event is mapped to the latent variable.
The latent variable as a function of $T_{\rm ap}$ and $E_{\rm ex}/A$ shows a sigmoid pattern, indicating that the trained autoencoder network treats the low and high temperature~(or low and high excitation energy) regions differently.
The area in the midst of the two phases represents those liquid-gas coexistence events that enter the spinodal region and are affected by the spinodal instability.

A confusion scheme combining the supervised and unsupervised learning has been adopted to obtain the limit temperature of the nuclear liquid-gas phase transition.
In the confusion scheme, the neural network is trained with data that are deliberately labelled incorrectly according to a proposed critical point, and the phase transition properties can be deduced from the performance curve, i.e., the total test accuracy as a function of the proposed critical point, of the neural network~\cite{NieNtP13}.
The total test accuracy reaches its minimum at $T_{\rm ap}'$ $\approx$ $T_{\rm lim}$.
The limiting temperature through the confusion scheme is $9.24\pm0.04~\rm MeV$, which is consistent with the $9.0\pm0.4~\rm MeV$ obtained from the traditional analysis of caloric curve~\cite{WadaPRC99}.

{\it{Crossover or first order phase transition}}
In general, as mentioned in the Introduction, the challenge being associated with high energy nuclear collision studies essentially can be viewed as an inverse problem. That is, assuming that all related physical factors (e.g., initial condition/fluctuations, QGP bulk properties, transport coefficients, freeze-out parameter, hadronic interactions, etc.) are given, then well-established theoretical models (e.g., relativistic viscous hydrodynamics with hadronic transport simulation) can be adopted to simulate the heavy ion collision process to give their final state observables, such a forward process is well understood. However, given instead only limited measurements on final state of heavy ion collisions, it's unclear how to disentangle those different influencing physical factors to decode back those corresponding early time dynamics. For high energy heavy ion collisions, two strategies exist in tackling this inverse problem with the help of statistical methods and machine learning, one is Bayesian inference with the task of parameter estimation for calibrating the chosen model(e.g., in Ref.~\cite{OmanaKuttan:2022aml} and those introduced in above texts), the other one is supervised machine learning to capture directly the inverse mapping from the final state to the corresponding physics of interest.

\begin{figure*}[htbp!]
    \centering
    \includegraphics[width = 0.9\textwidth]{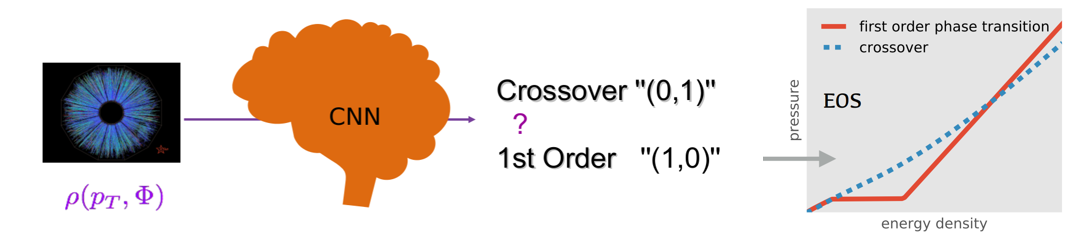}
    \caption{A schematic plot for QCD transition classification with HICs final particle spectra.}
    \label{fig:cnn_eos}
\end{figure*}

In Ref.~\cite{Pang:2016vdc} it is proposed to use deep convolutional neural network to capture the direct inverse mapping from the final state information to the cared early time happened QCD transition types. 
This is inspired by the success of image recognition in computer vision, that although the inverse mapping might be very implicit, but it's possible to use deep neural networks to decode and represent it in the sense of big data supervisedly. 
The needed training data can be prepared from well-established model simulation for heavy ion collisions, such as the state of the art 3+1-dimensional viscous hydrodynamics~\cite{Pang:2012he,Shen:2014vra}, where diversity can be introduced in by varying these different physical factors (i.e., parameters within the simulation). 
As an exploratory study a binary classification task is targeted, where the Deep CNN is trained to identify the QCD transition type embedded within the collision dynamics to be crossover or first order solely based upon the final pion spectra $\rho(p_T,\phi)$, as shown in Fig.~\ref{fig:cnn_eos}.
Basically, the EoS of the hot and dense matter enters the hydrodynamic simulations as a crucial ingredient, embedded in which the nature of QCD transition (first order or crossover) can affect strongly the hydrodynamic evolution due to the shape of pressure gradient. 
As the input to the deep CNN, final charged pion's spectra at mid-rapidity are obtained from the Cooper-Frye formula in each hydrodynamic simulation,
\begin{align}
\rho(p_T,\phi)=\frac{dN_i}{dY p_Tdp_Td\phi}=g_i\int_\sigma p^{\mu}d\sigma_{\mu}f_i(p\cdot u)
\end{align}
with $N_i$ the particle number density, $Y$ the rapidity, $g_i$ the degeneracy, $d\sigma_\mu$ the freeze-out hypersurface element and $f_i$ the thermal distribution. The training dataset of $\rho(p_T,\phi)$ is generated from event-by-event hydrodynamic package CLVisc~\cite{Pang:2012he} with fluctuating AMPT initial conditions, with which supervised learning using CNN is performed for binary classification in identifying the QCD transition types.

\begin{figure}[htb]
\includegraphics[width=\columnwidth]{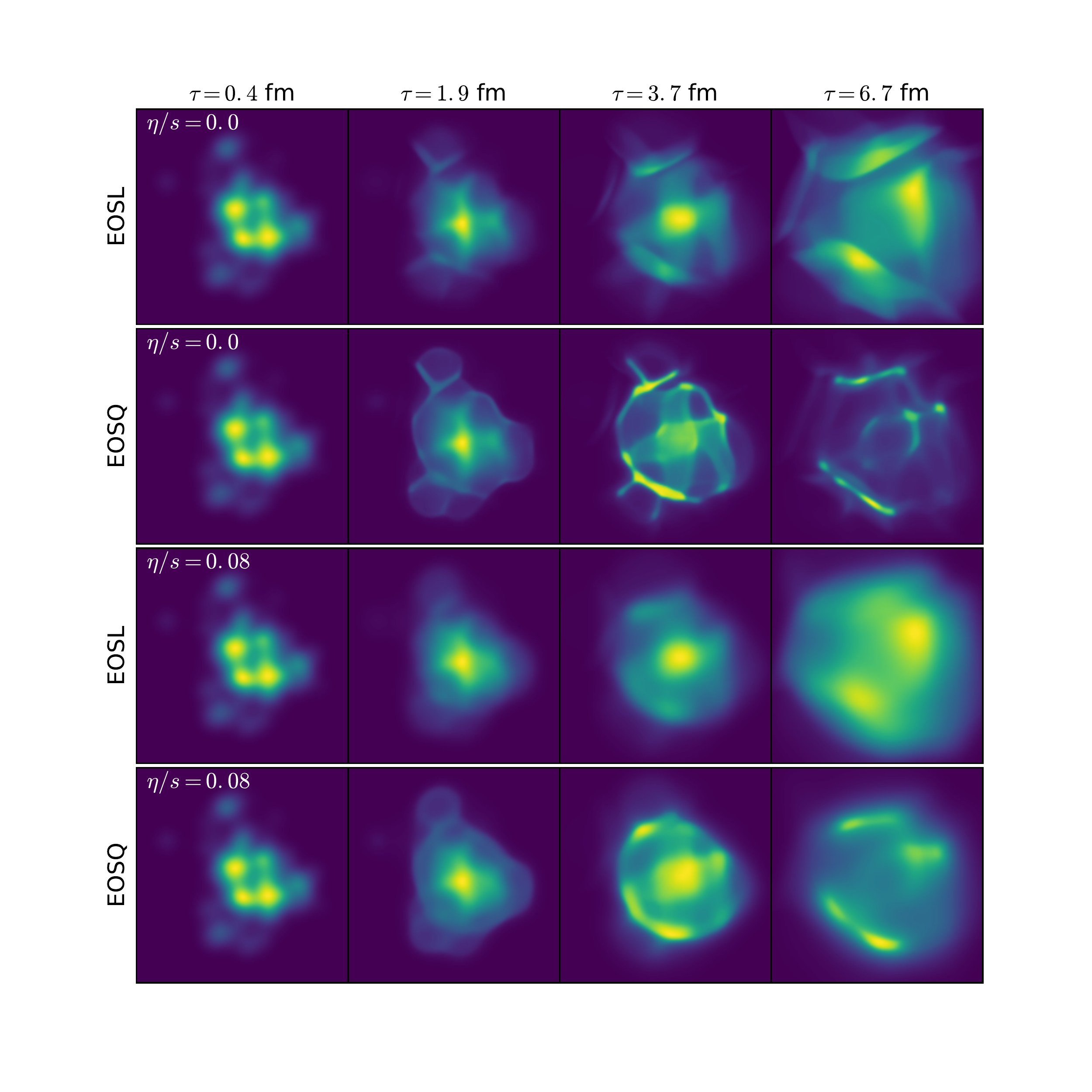}
\caption{The evolution history of quark gluon plasma simulated by relativistic hydrodynamic model CLVisc, starting from the same initial condition with four different parameter combinations. From top to down, each row represents four snapshots at different time, using different combinations of equation of state and shear viscosity over entropy density ratio, where EOSL represents lattice QCD equaiton of state with a crossover transition between QGP and hadron resonance gas, EOSQ represents an equation of state with a first order phase transition between QGP and hadron resonance gas. \label{fig:crossover_vs_1st}}
\end{figure}
Shown in Fig.~\ref{fig:crossover_vs_1st} is the space-time evolution histories of QGP expansion starting from the same initial condition with different conditions, in relativistic hydrodynamic simulations using CLVisc. For EOSQ with a first order phase transition, the pressure gradient is zero in the mixed phase. Multiple ridge structures are formed with a first order phase transition in the EoS because the expansion of QGP is mainly driven by the pressure gradient and the acceleration is 0 in the mixed phase. On the other hand, the expansion histories are quite different
when the shear viscosity is not 0. Different evolution histories lead to different final state particle spectra in momentum space. 

To verify the robustness of the trained Deep CNN in this QCD EoS recognition task, the testing set was simulated from a different hydrodynamics package, or with different initial fluctuating conditions (IP-Glasma or MC-Glauber) and different $\eta/s$ parameters. The conventional observables like elliptic flows $v_2$ and integrated particle spectrum are shown to be incapable of distinguishing the two QCD transition classes on these testing set, while the trained deep CNN gives on average $95\%$ classification accuracy which is robust against other contamination factors, especially from initial fluctuations and shear viscosity. For comparison, the classification accuracy from traditional machine learning algorithms such as decision tree, random forest, support vector machine or gradient boosting is to the best approximately $80\%$. The good performance of the trained deep CNN first indicate that imprint of the early time transition dynamics is not fully washed out by the collision evolution, and is still encoded inside the final state information. And, the inverse mapping from final state observables to the QCD transition information can be well captured by the deep CNN from supervised training strategy, thus enabling discriminative and traceable encoder for the dynamical information of QCD transition. In this way, the constructed deep CNN act as an “EoS-meter" to efficiently bridge the heavy ion collision experiments to QCD bulk matter physics. This study paves a path to the success of experimental study on QCD EoS and the search for the critical end point of QCD phase diagram. It should be noted that, this study didn't consider the afterburner hadronic cascade effects, thus the conclusion about the direct inverse mapping is made on the level of pure hydrodynamic evolution.

Later this strategy was further deepened in a series of works for more realistic scenarios, like to take into account the afterburner hadronic cascade via incorporating UrQMD following the hydrodynamics evolution~\cite{Du:2019civ,Du:2020poe}, to consider non-equilibrium dynamics of phase transition's influence e.g., spinodal decomposition~\cite{Steinheimer:2019iso,Steinheimer:2021hoc} or Langevian dynamics~\cite{Jiang:2021gsw}, to further also include the more realistic experimental detector effects through detector simulation with Hits or Tracks as input~\cite{OmanaKuttan:2020brq,OmanaKuttan:2020btb}, perform unsupervised outlier detection in heavy ion collisions~\cite{Thaprasop:2020mzp} and identifying nuclear symmetry energy~\cite{Wang:2021xbb}. Specifically, in Ref.~\cite{OmanaKuttan:2020btb} it was shown that using just the detector output directly, PointNet models can be used to classify collision events simulated by an EoS associated with a first order phase transition from an EoS with a crossover transition. The PointNet models take the reconstructed tracks from the CBM detector (simulated with CBMRoot) following by the hybrid- UrQMD events. It achieves a binary classification accuracy around $96\%$ when trained on collision events for impact parameter ranging from 0 to 7 $fm$. When the model training set was shrunk to mid-central region with $b=0\sim3 fm$, the model performance increased with an accuracy about $99\%$. A combination of training set from both peripheral and mid-central collisions also works and result in a classifier being able to identify the phase transition type across different centralities meanwhile not compromising the accuracy for central region.

{\it{Active learning for QCD EoS}}
First principle calculations using Lattice QCD provide the equation of state of hot nuclear matter at high temperature and zero baryon chemical potential. Because of fermion's sign problem, Lattice QCD fails to compute the nuclear EoS at finite $\mu_B$ at present. Using Taylor expansion, it is possible to get the nuclear EoS at small $\mu_B$ that is close to zero, approximately. The BEST collaboration constructed one nuclear EoS with critical end point by mapping the 3D Ising model with the Tylor expansion result. However, the model contains 4 free parameters whose values determine the size and location of the critical end point. Some combinations of these parameters lead to unphysical EoS, e.g., acausal or unstable.

Supervised learning can help to mapping out unphysical regions of parameter combinations. However, labelling is computationally expansive in this task. For thermaldynamic stability, one has to check the positivity of energy density, pressure, entropy density, baryon density, second order baryon susceptibility $\chi_2^B$ and the heat capacity $(\partial S / \partial T)_{n_B}$, as well as the causality condition,
\begin{align}
0 \le c_s^2 \le 1
\end{align}
where $c_s$ is the speed of sound of hot nuclear matter.  

The authors use active learning to find the most informative parameter combinations before labelling them \cite{Mroczek:2022oga}. In active learning, the network is first trained using a small amount of labelled data. Then the trained network is employed to make predictions on all samples from a large unsupervised pool. If the network is uncertain about one parameter combination, e.g., it predicts that this group of parameter combination will lead to a EoS that is unphysical with probability 51\%, then this sample lives on the decision boundary and should be quite informative and important for the network. Labelling this sample will improve the performance of the network more than labelling easy samples. The newly labelled sample will be moved out from the pool and will take part in supervised learning later. 

\begin{figure*}[t]
	\centerline{\includegraphics[width=1.0\linewidth]{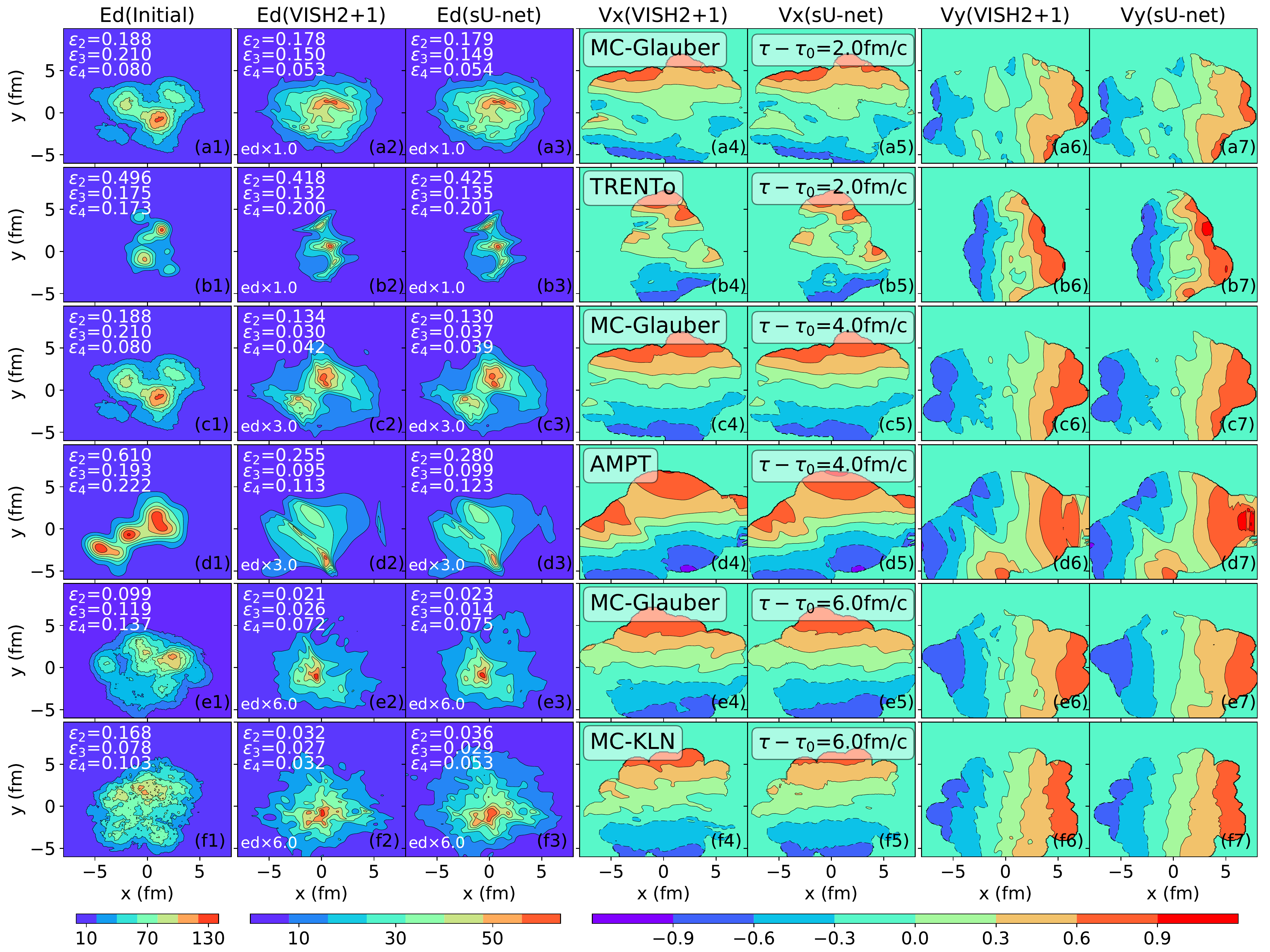}}
	\caption{ Energy density and flow velocity profiles, predicted by {sU-Net} and calculated from {VISH2+1} for six test initial profiles of {MC-Glauber}, {MC-KLN}, {AMPT} and {TRENTo}~\cite{Huang:2018fzn,Huang:2018rtk}.}
	\label{cmp}
\end{figure*}

{\it{Deep learning accelearate Relativistic Hydrodynamics}}
Relativistic Hydrodynamics is a powerful tool to simulate the QGP expansion and study the flow observables in
Relativistic Heavy Ion Collisions at RHIC and the LHC energies~\cite{Teaney:2009qa,Romatschke:2009im,Heinz:2013th,Gale:2013da,Song:2013gia,Song:2017wtw}. For ideal hydrodynamics with zero net charge densities, it solves the transport equations of the energy momentum tensor:
\begin{equation}
\partial_{\mu}T^{\mu\nu}=0
\end{equation}
where $T^{\mu\nu}=(e+p) u^{\mu}u^{\nu}-p g^{\mu\nu}$,
$e$ is the energy density, $p$ is the pressure, and $u^\mu$ is the four velocity. For traditional simulations,  hydrodynamics numerically solve these transport equations with some particular Algorithm such as SHASTA, LCPFCT, which translate the initial conditions to final state profiles through its non-linear evolutions~\cite{Kolb:2003dz,Song:2009gc,Heinz:2013th,Song:2017wtw}.

In the recent work~\cite{Huang:2018fzn,Huang:2018rtk}, a deep neural network, called  stacked U-net (sU-net), is designed and trained to learn the initial and final state mapping from the nonlinear hydrodynamic evolution. The constructed sU-net belongs to the architecture of encoder-decoder network, which contains four blocks of U-net with residual connections between them. For each U-net, there are three convolutional and deconvolutional layers with Leaky ReLU and softplus activation functions employed for the inner and output layers. By concatenating the feature maps along the channel dimension, the output of the first two convolution layers is fed into the last two deconvolution layers. For more details, please refer to~\cite{Huang:2018fzn,Huang:2018rtk}.

The training and test data (the profiles of initial and final energy momentum tensor $T^{\tau\tau}$, $T^{\tau x}$, $T^{\tau y}$) are generated from {VISH2+1} hydrodynamics~\cite{Song:2007ux,Song:2007fn} with zero viscosity, zero net baryon density and longitudinal boost invariance. In more detail, sU-net is trained with  10000 initial and final profiles from VISH2+1 with MC-Glauber initial conditions~\cite{Miller:2007ri,Hirano:2009ah}, and then tested for its  predictive power using the profiles of four different types of initial conditions, including MC-Glauber~\cite{Miller:2007ri,Hirano:2009ah},  MC-KLN~\cite{Hirano:2009ah,Drescher:2006ca}, AMPT~\cite{Pang:2012he,Xu:2016hmp,Zhao:2017yhj} and TRENTo~\cite{Moreland:2014oya}. Fig.~\ref{cmp} shows the final energy density and flow velocity predicted by sU-net, together with a comparison with the hydrodynamic results. It can be seen that the trained sU-net can capture the magnitudes and structures of both energy density and flow velocity. In particular, panels (b), (d) and (f) show that the network, trained with data sets from MC-Glauber initial conditions, is also capable of predicting the final profiles of other types of initial conditions. Ref.~\cite{Huang:2018fzn,Huang:2018rtk} also calculated the eccentricity coefficients which evaluate the deformation and inhomogeneity of a large number of the energy density profiles, which demonstrated that the predictions from sU-net almost overlap with the results from VISH2+1.

Compared with the 10$\sim$20 minutes simulation time of VISH2+1 on a traditional CPU, sU-net takes several seconds to directly generate the final profiles for different types of initial conditions on one P40 GPU, which greatly accelerates the traditional hydrodynamic simulations. However, the designed and trained sU-net in Ref.~\cite{Huang:2018fzn,Huang:2018rtk} mainly focuses on mimicking the 2+1-dimensional hydrodynamic evolution with fixed evolution time. For more realistic implementation, it is also important to explore the possibilities of mapping the initial profiles to the final profiles of the particles emitted on the freeze-out surface  of the relativistic heavy ion collisions.

\section{In-medium effects}
\label{sec:med_eff}
\textit{spectral function reconstruction} Accessing Real-time properties of QCD (or a many-body system in general) remain a notoriously difficult problem, because the non-perturbative computations, such as lattice field simulations or functional methods, usually operate in Euclidean space-time (after a Wick rotation $t\to i t\equiv\tau$) and thus can only provide Euclidean correlators (i.e., in imaginary time). Then the analytic continuation of these discrete noisy data is often ill-posed. But quantitatively understanding the real-time dynamics determined by the Minkowski correlator is important and interesting in itself, for example for understanding scattering processes, transport or non-equilibrium phenomena that happen in heavy ion collisions. The Minkowski correlator is usually accessed from the Euclidean correlator via spectral reconstruction. 

The associated ill-posed problem can be cast as a Fredholm equation of the first kind, 
\begin{align}
g(t)=\int_a^b K(t,s)\rho(s)ds, \label{eq:fredholm}
\end{align}
with the goal of retrieving the function $\rho(s)$ given the kernel function $K(t, s)$ but limited information about $g(t)$. It has been well shown that the required inverse transform becomes ill-conditioned once only a finite set of data points with non-vanishing uncertainty are available for $g(t)$. In the context of QFT, one could simply approach this problem via the Kaellen-Lehmann spectral representation of the correlators, thus taking the kernel function to be 
\begin{align}
K(t,s)=s(s^2+t^2)^{-1}\pi^{-1}, \label{eq:KL_kernel}
\end{align}
where the $\rho(s)$ involved are usually called spectral functions. The task of reconstructing the spectral function from the correlator measurements (from the lattice calculation) needs to be regularized to make sense of the inverse problem involved. Over the past few decades, many different regularization techniques have been explored for this ill-conditioned inverse problem, such as Tikhonov regularization, maximum entropy methods, Bayesian inference techniques.

Recently, deep learning based strategies have also been explored to tackle spectral reconstruction, which can be mainly categorized into two schemes: data-driven supervised learning approaches and unsupervised learning based approaches.
The first application of domain-knowledge-free deep learning methods to this ill-conditioned spectral reconstruction (also called analytic continuation) appears in Ref.~\cite{2018PhRvB..98x5101Y} in the context of general quantum many-body physics, which shows the good performance of deep neural networks via supervised training in the cases of a Mott-Hubbard insulator and also metallic spectrum. In particular, the convolutional neural network was found to achieve better reconstruction than the fully connected network, with performance superior to the MEM, one of the most widely used conventional methods.
In Ref.~\cite{PhysRevLett.124.056401} the authors adopted a similar strategy, but also introduced Principal Component Analysis (PCA) to reduce the dimensionality of the QMC simulated imaginary time correlation function of the position operator for a harmonic oscillator linearly coupled to an ideal heat bath. 

Ref.~\cite{Kades:2019wtd} also from a data-driven perspective, pushed a similar strategy to spectral function reconstruction into the QFT context and considered the K\"allen--Lehmann (KL) spectral representation as the accessible propagator, $G(p)=\int^{\infty}_0\frac{d\omega}{\pi}\frac{\omega\rho(\omega)}{\omega^2 + p^2}$, which basically takes the kernel in the Fredholm equation as the K\"allen--Lehmann(KL) kernel. For the dummy spectral functions, the superposition of Breit-Wigner peaks was used, based on the perturbative one-loop QFT-derived parameterisation $\rho^{BW}(\omega)=4A\Gamma\omega/((M^2+\Gamma^2-\omega^2)^2+4\Gamma^2\omega^2)$. Two types of deep neural networks have been studied, both with the noisy propagator as input, but with different outputs: one estimates the parameters (e.g. $\Gamma_i$, $M_i$ for the collection of Breit-Wigner peaks) of the spectral function (denoted PaNet), the other tries to reconstruct directly the discretized data points of the spectral function (denoted PoNet).


\begin{figure*}[htbp!]
    \centering
    \includegraphics[width = 0.9\textwidth]{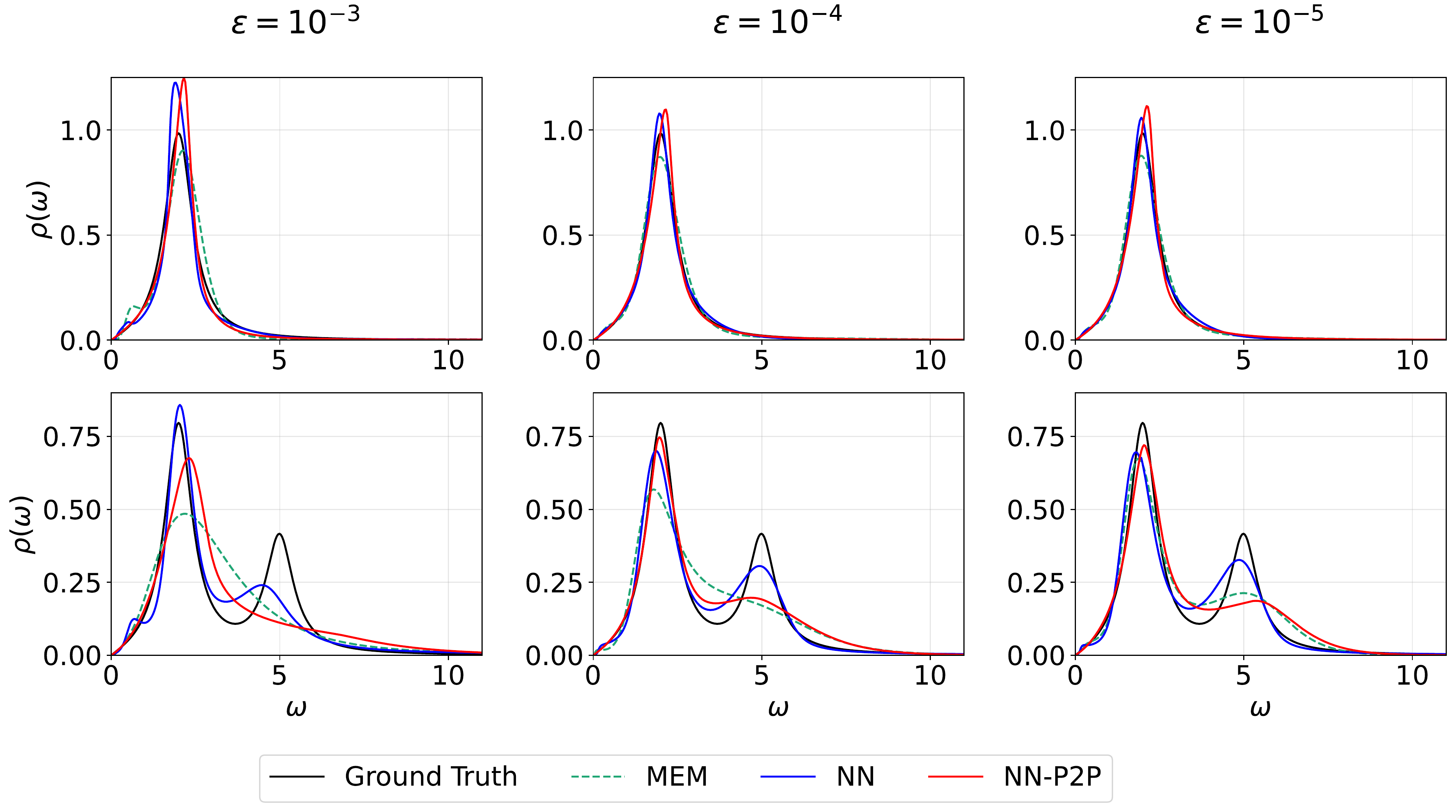}
    \caption{The spectral functions reconstructed from MEM, NN and NN-P2P, under different Gaussian noises added to the propagator data with $N_p = 25$, and $N_\omega = 500$ for the spectral. Taken from Ref.~\cite{Wang:2021jou}.
    \label{fig:rec}}
\end{figure*}

As another non-parametric representation, Gaussian Processes (GP) are used in the reconstruction of the 2+1 flavor QCD ghost and gluon spectral function in Ref.~\cite{Horak:2021syv}. In general, the GP can define a probability distribution over families of functions, typically characterized by the chosen kernel function. In Ref.~\cite{Horak:2021syv} the GP is assumed to describe the spectral function,
\begin{align}
\rho(\omega)\sim \mathcal{GP}(\mu(\omega),C(\omega, \omega')),
\end{align}
with $\mu(\omega)$ the mean function often set to zeros, and $C(\omega,\omega')$ the covariance dictated by the kernel function used, to which a common standard choice is the radial basis function (RBF) kernel,
\begin{align}
C(\omega,\omega')=\sigma_C e^{-\frac{(\omega-\omega')^2}{2l^2}},
\end{align}
including tunable hyperparameters $\sigma_C$ for overall magnitude and $l$ a length scale. Then this GP represented prior can be plugged into the Bayesian inference procedure with lattice data for the ghost dressing function and gluon propagator for constructing the likelihood. Ref.~\cite{Horak:2021syv} also specifically extended the lattice data, e.g., for ghost they extended the dressing function into the deep IR and also constrain the low-frequency behavior by spectral DSE results~\cite{Horak:2021pfr}, and for gluon it's extended into the UV with previous fRG computation results~\cite{Cyrol:2018xeq}. This reduces the variance in the solution space and enhanced the stability compared to the inference without such extensions. It's shown that while approximately fulfilling the Oehme-Zimmermann superconvengence (OZS) condition for gluon, the reconstruction with GP regression in this work accurately reproduces the lattice data within the uncertainties with deviations for gluon propagator stronger in some region than for the ghost dressing function. For the spectral function, the reconstruction shows similar peak structure to previous fRG reconstruction of the Yang-Mills propagator~\cite{Cyrol:2018xeq}.

In Ref.~\cite{Wang:2021jou, Wang:2021cqw, Shi:2022yqw}, the authors devised an unsupervised approach based on deep neural network (DNN) representation for the spectral function together with automatic differentiation (AD) to reconstruct the spectral, which does not need training data preparation for supervision (Note that similar DNN based inverse problem solving strategy within AD framework is also used to reconstruct Neutron Star EoS from astrophysical observables~\cite{Soma:2022qnv, Soma:2022vbb} and inferring parton distribution function of pion in lattice QCD studies~\cite{Gao:2022iex}). The introduced DNN representation can preserve the smoothness of the spectral function automatically, thus naturally help to regularize the degeneracy issue in this inverse problem, because it's analyzed in Ref.~\cite{Shi:2022yqw} that the degeneracy is related to the null-modes of the investigated kernel function, which will usually induce oscillation for the reconstructed spectral function. Specifically, the DNN represented spectral, $\vec{\rho}=[\rho_1,\rho_2,...,\rho_{N_{\omega}}]$, can be converted into the propagator under the discretization scheme as $D(p)=
\sum^{N_{\omega}}_i K(p,\omega_i)\rho_i\Delta\omega$. Then the loss function over the propagator as compared to lattice data, $\mathcal{L}=\sum^{N_p}_i(D_i-D(p_i))^2/\sigma_i$, can be evaluated and provide guide for the tuning over DNN represented spectral. Taking gradient-based algorithms, the derivative of the loss with respect to network parameters can be derived as
\begin{align}
\nabla_{\theta}\mathcal{L}=\sum_{j,k}K(p_j,\omega_k)\frac{\partial\mathcal{L}}{\partial D(p_j)}\nabla_{\theta}\rho_k. \label{eq:ad1}
\end{align}
with $\nabla_{\theta}\rho_k$ computed easily under standard backward propagation for the network. 

For the DNN representation of spectral, two different schemes were investigated in this work: one used the multiple outputs of an L-layers neural network to represent in list format the spectral function (denoted as NN), and the other directly use a feedforward neural network as parameterization (denoted as NN-P2P) over the spectral as a function of frequency, $\rho(\omega)$. For the training, the Adam optimizer is adopted, and the $L_2$ regularization is set in the warm-up beginning stage under an annealing strategy until a small enough regularization strength value (set as smaller than $10^{-8}$ in the calculation), this can relax the regularization to obtain a hyperparameter independent inference results. For the direct NN list representation, a quenched implementation of smoothness condition $\lambda_s\sum^{N_{\omega}}_{i=1}(\rho_i - \rho_{i-1})^2$ is also performed with $\lambda_s$ decreased from $10^{-2}$ to 0. Such newly devised unsupervised spectral reconstruction method got approved for the uniqueness of the solutions with manifestation both analytically and numerically~\cite{Shi:2022yqw}.
As Fig.~\ref{fig:rec} shows, on superposed Breit-Wigner peaks this method was demonstrated to outperform the traditional MEM method, especially on multi-peaks spectra with large measurement noise situations. 

\begin{figure*}
  \begin{center}
    \includegraphics[width=0.8\textwidth]{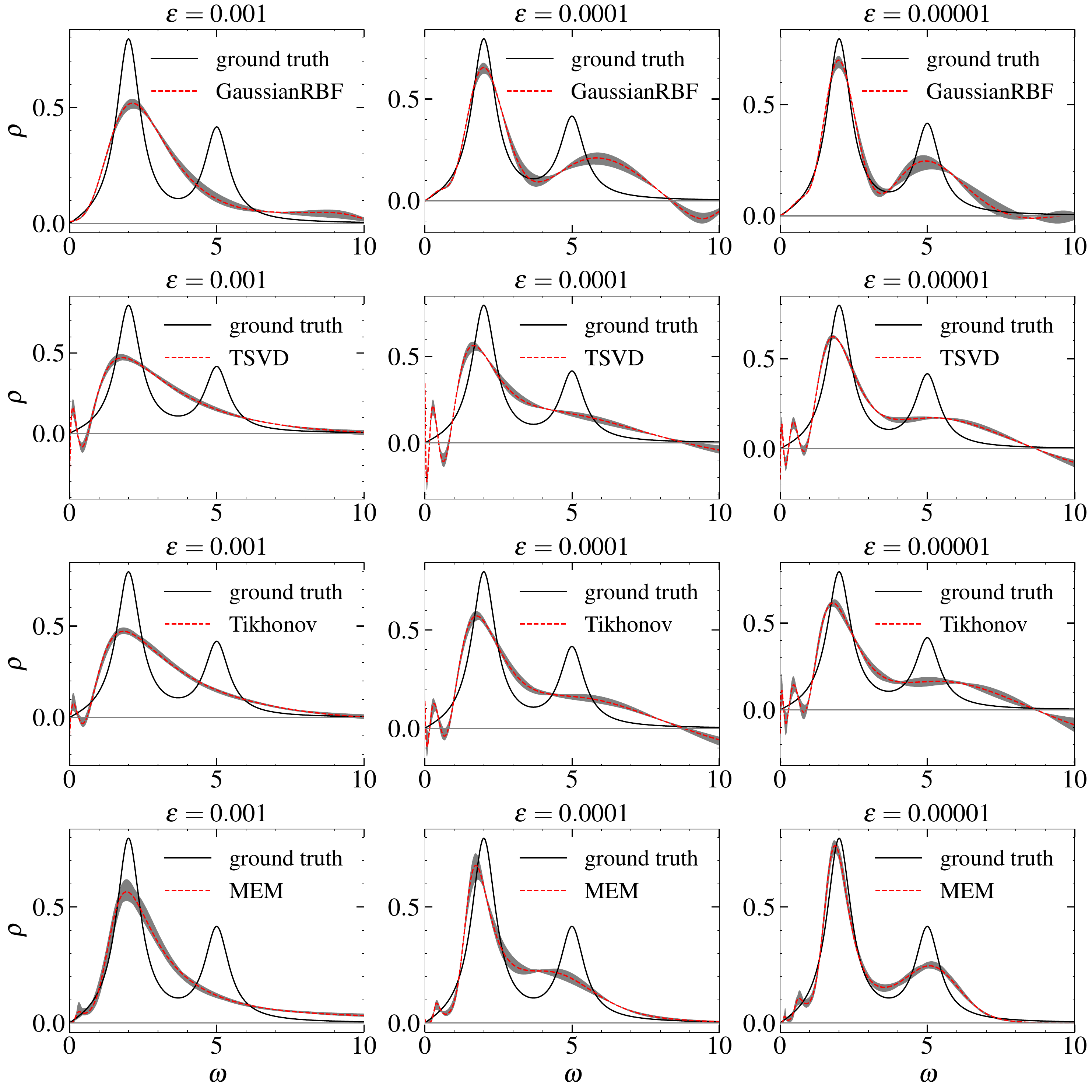}
    \caption{The constructed spectral functions obtained from RBFN, TSVD, Tikhonov and MEM, using the correlation data generated by the mock SPF of mixing two Breit-Wigner distributions. For left to right panels, different Gaussian noises are added to the correlation data with $\epsilon = 0.001$,  $0.0001$ and  $0.00001$~\cite{Zhou:2021bvw}.}
    \label{fig:RBFNoise}
  \end{center}
  \end{figure*}

\begin{figure*}[!hbtp]
\centering
\includegraphics[width=0.33\textwidth]{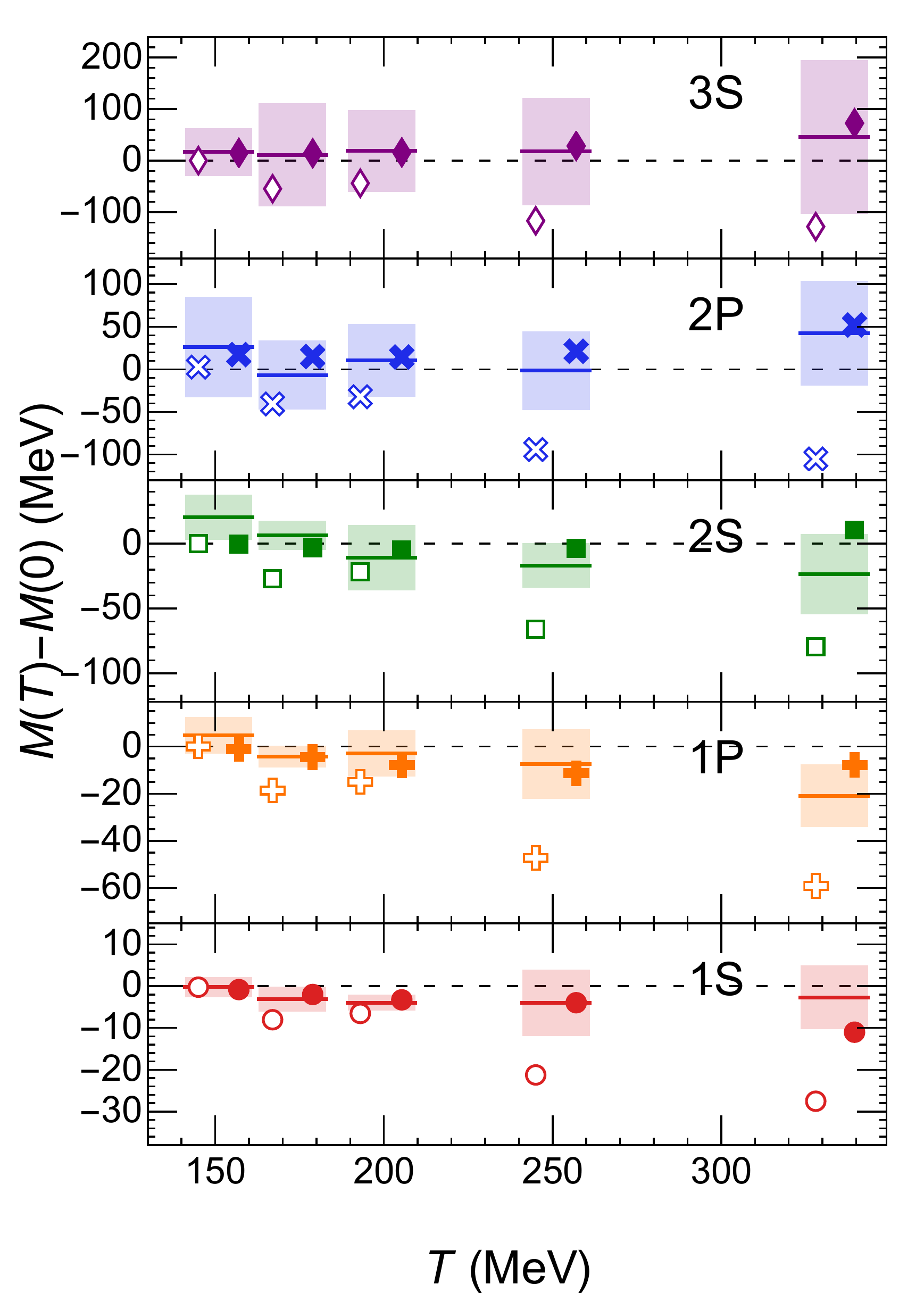} 
\includegraphics[width=0.33\textwidth]{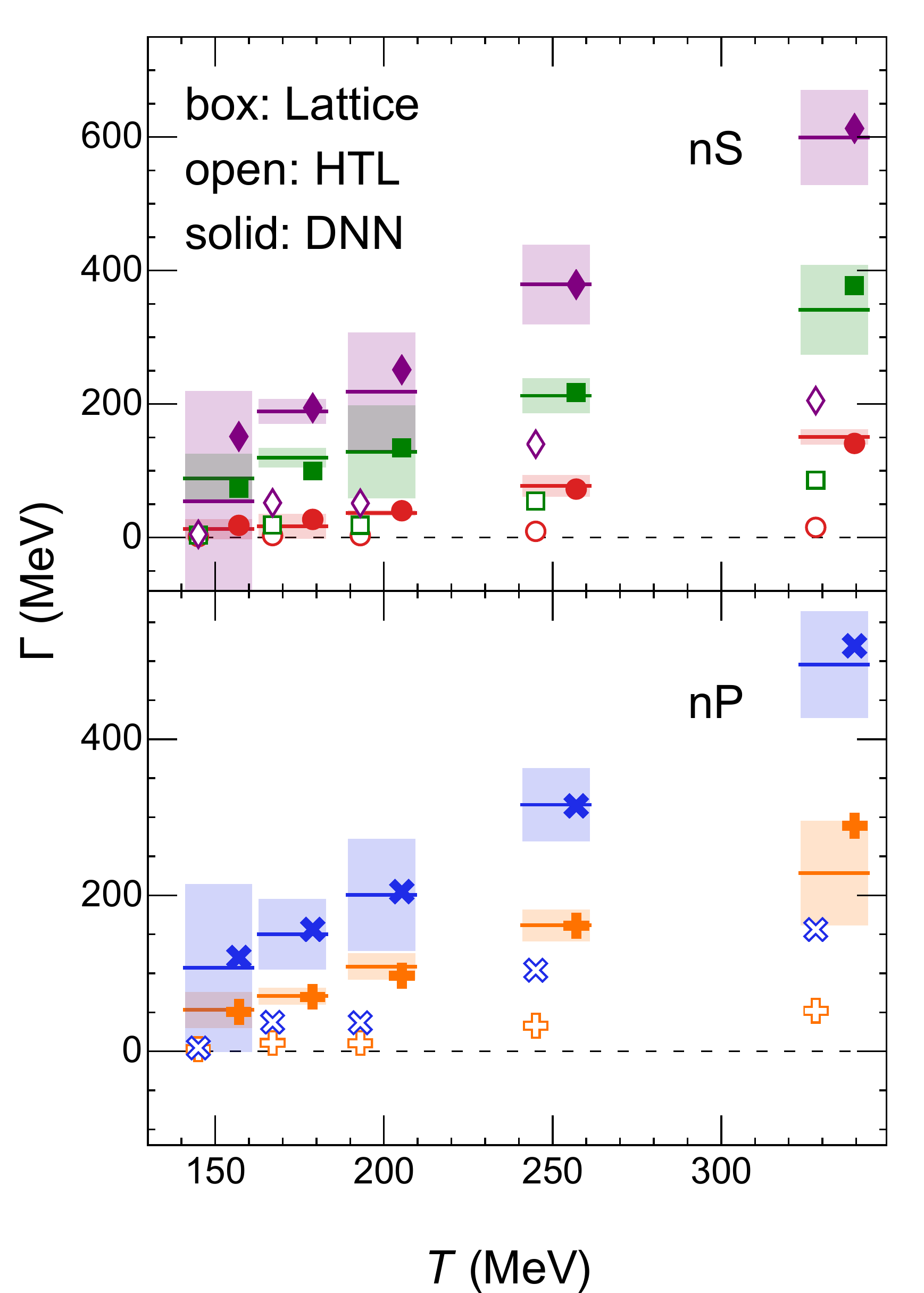}
\includegraphics[width=0.33\textwidth]{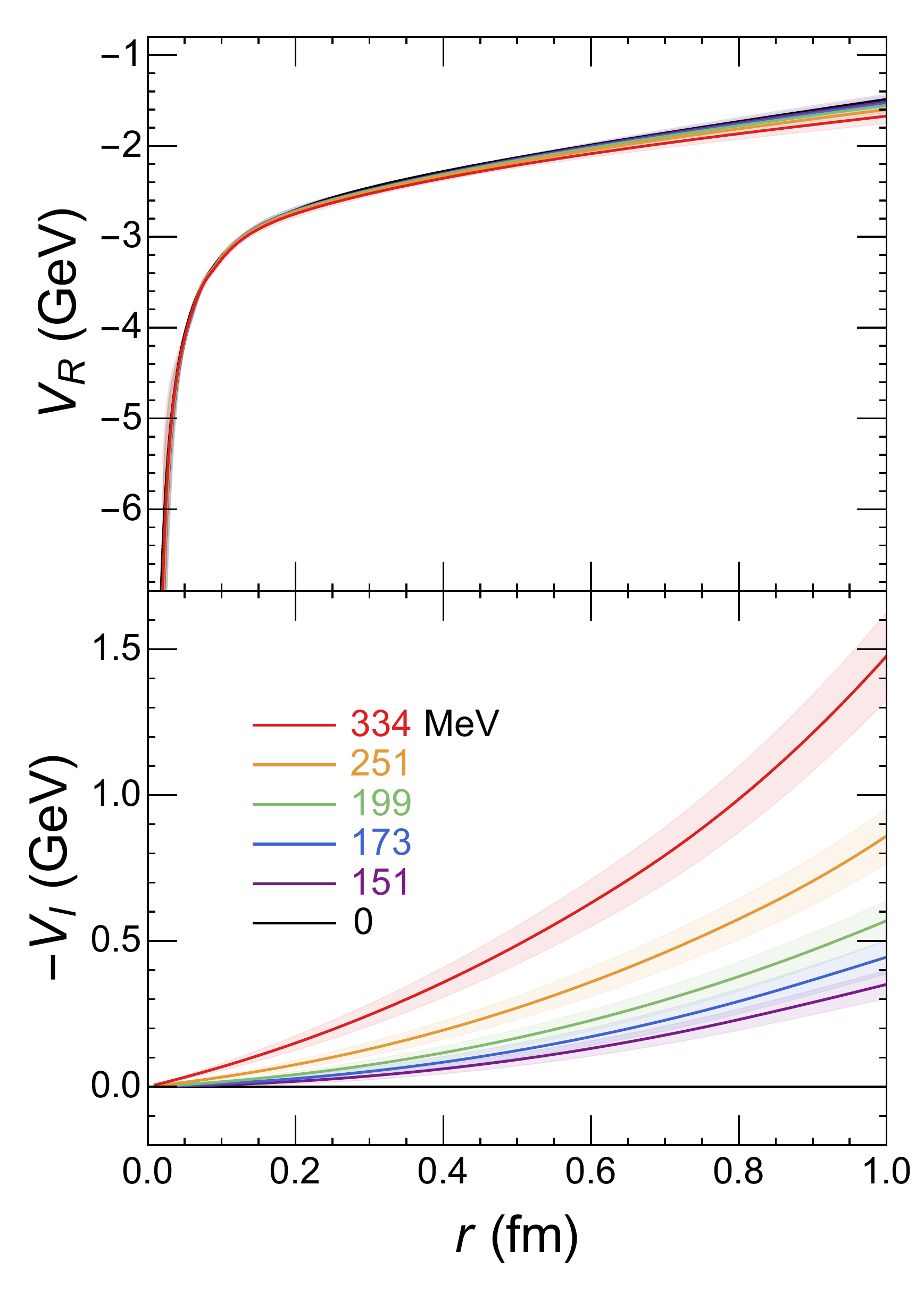}
\caption{Piciture taken from Ref.~\cite{Shi:2021qri}. Left and Middle: In-medium mass shifts with respect to the vacuum mass (left) and the thermal widths (right) of different bottomonium states obtained from fits to LQCD results of Ref.~\cite{Larsen:2019zqv} (lines and shaded bands) using weak-coupling motivated functional forms~\cite{Lafferty:2019jpr} (open symbols) and DNN based optimization (solid symbols). The points are shifted horizontally for better visualization. 
$\Upsilon(1S)$, $\chi_{b_0}(1P)$, $\Upsilon(2S)$, $\chi_{b_0}(2P)$ and $\Upsilon(3S)$ states are represented by red circles, orange pluses, green squares, blue crosses and purple diamonds, respectively. Right: The DNN reconstructed real (top) and imaginary (bottom) parts of the heavy quark potential at temperatures $T = 0$(black), $151$(purple), $173$(blue), $199$(green), $251$(orange), and $334$~MeV(red). The uncertainty bands represent the $68\%(1\sigma)$ confident region.
\label{fig.mass}}
\end{figure*}

Beside further Gaussian-like and Lorentian-like spectral reconstruction tests, the newly devised framework in Ref.~\cite{Wang:2021jou, Wang:2021cqw} was also validated in two physics motivated tests, one is for non-positive definite spectral reconstruction, which is beyond classical MEM applicability scope but often encountered for spectral functions related to confinement phenomenon to e.g., gluons and ghosts, or
thermal excitations with long-range correlation in strongly
coupled system; the other one is for hadron spectral function encoded in temperature dependent thermal correlator with lattice QCD noise-level noises. On both of these two further physical cases this proposed DNN and automatic differentiation based method with NN representation works consistently well while traditional MEM based methods loses the peak information or fail in confronting the non-positiveness.

The spectral function can also be reconstructed from finite correlation data through implementing
the radial basis functions networks (RBFN), which is a multilayered perceptron model based on the radial basis functions(RBF)~\cite{broomhead1988radial,schwenker2001three}. RBFN has been widely used in feature extraction, classification, regression and so on~\cite{beheim2004new,wang2002point,carr1997surface,chen2018deep}. In Ref.~\cite{Zhou:2021bvw}, the spectral function
$\rho\l(\omega\r)$ was first approximately described by a linear combination of RBFs:
\begin{equation}
  \rho\l(\omega\r)=\sum_{j=1}^N w_{j}\phi\l(\omega-m_j\r),\label{eq:linear_summation}
\end{equation}
where $\phi$ is the active RBF with an adjustable weight $w_{j}$  and  an adjustable center $m_j$, which
can take a Gaussian form $\phi(r)=e^{-\frac{r^2}{2a^2}}$  or a MQ form
$\phi(r)=(r^2+a^2)^{\frac{1}{2}}$, etc. Here, $a$ is the shape parameter, which is adjustable and essential for the regularization. Then, the inverse mapping problem of constructing the spectral function is transformed into calculating  the linear weights of RBF, which naturally enables a smooth and continuous reconstruction.

To calculate these parameters in Eq.(\ref{eq:linear_summation}), Ref.~\cite{Zhou:2021bvw} constructed a neutral network called radial basis function networks (RBFN), which is a three-layers feed-forward neural network with the active RBFs built in the hidden layer. After discrete the spectral function, Eq.(\ref{eq:linear_summation}) is transformed  into a matrix form $\l[\rho\r]=\l[\Phi\r]\l[W\r]$. Then, the correlation functions in Euclidean space with the integral spectral representation $G\l(\tau,T\r)=\int^\infty_0 \frac{d\omega}{2\pi}\rho\l(\omega ,T\r)K\l(\omega,\tau,T\r)$ becomes a matrix form:
\begin{equation}\label{eq:RBFMatForm}
  G_i=\sum_{j=1}^{M} \sum_{k=1}^N K_{ij}\Phi_{jk}w_k\equiv \sum_{k=1}^{M} \tilde{K}_{ik}w_k, \ \ \ i=1...\widehat{N}
\end{equation}
where $\tilde{K}$ is a $\widehat{N}\times M $ matrix associated with the integration kernel, $\widehat{N}$ is the number of data points for the correlation function $G_i$. Note that the spectral function $\rho(\omega_i)$ has been discretized into N parts with $m_i=\omega_i,\ i=1...N$  and M is set to $M=N=500$. To obtain $w_j$,
one could further implement the TSVD method or a deep neural network~\cite{Zhou:2021bvw}.  Compared with other machine learning approach based on the supervised learning~\cite{yoon2018analytic, fournier2020artificial, Kades:2019wtd}, this method can be rapidly trained and free from the over-fitting problem.

Fig.~\ref{fig:RBFNoise} shows a comparison of  the constructed spectral functions from RBFN, TSVD, Tikhonov and MEM, using the correlation data generated by a mock SPF. Such mock SPF is obtained by mixing two Breit-Wigner distributions:
$\rho_{Mock}(\omega)=\rho_{BW}(A_1,\Gamma_1, M_1, \omega)+\rho_{BW}(A_2,\Gamma_2, M_2, \omega)$ with $\rho_{BW}(A_i,\Gamma_i, M_i, \omega)=\frac{4A_i\Gamma_i \omega}{\l(M_i^2+\Gamma_i^2-\omega^2\r)^2+4\Gamma_i^2 \omega^2}$. The parameters for the mock SPF in Fig.~\ref{fig:RBFNoise} are set to $A_1=0.8, M_1=2, \Gamma_1=0.5; A_2=1, M_2=5, \Gamma_2=0.5$. Here, one generates  $30$  discrete correlation data using the
Euclidean correlation functions of the mock SPF, together with a noise added, $G_{\text{noise}}(\tau_i) = G(\tau_i) + \text{noise}$.

Compared with the results from traditional methods, RBFN provides a better description of the spectra functions, especially for the low frequency part. It also almost reproduces the first peak of the mock SPF using the correlation data with smaller noise $\epsilon = 0.00001$. In contrast, Tikhonov, TSVD and MEM present some oscillation behavior at low frequency. For such a task of extracting the transport coefficients from the Kubo relation, an improved reconstruction of the spectra functions at low frequency is especially important. Although
RBFN fails to reconstruct the second peak of the mock SPF, it is the only method that reduces the oscillation at the low frequency, compared with the  three commonly used ones. Ref.~\cite{Zhou:2021bvw} also compared the Gaussian and MQ  RBF used in the network and  found that the Gaussian RBF gives better construction of the SPF, including the location and the width of the peak.  With the mock data generated from the spectral function of energy  momentum tensor,  Ref.~\cite{Zhou:2021bvw} also demonstrated that the RBFN  method  gives a precise and stable extraction of the transport coefficients.

{\it{in-medium heavy quark potential}} 
As an important probe for the properties of the created QGP in heavy ion collisions, heavy quarkonium (the bound state of heavy quark and its anti-quark) is intensively measured in experiments and studies theoretically~\cite{Zhou:2014kka, Zhao:2020jqu}, the investigation and calculation to which requires understanding to the in-medium heavy quark interaction. Basically, the heavy quarkonium provide a calibrated QCD force, since in vacuum the simple Cornell potential can reproduce well the spectroscopy of heavy quarkonium, and when we put the bound state into QCD medium the color screening effects would naturally happen and weaken the interactions between the heavy quarks, beyond which, a non-vanishing imaginary part manifested as thermal width is argued to show up according to both the one-loop hard thermal loop (HTL) perturbative QCD calculations~\cite{Laine:2006ns,Beraudo:2007ky} and the recent effective field theory studies e.g., pNRQCD~\cite{Brambilla:2008cx,Brambilla:2010vq}. However, the non perturbative treatment like lattice QCD is necessary because it's difficult to get satisfactory description of the strong interaction dictated in-medium heavy quarkonium solely from perturbative calculations. From those EFTs studies, it's also shown that a potential based picture can provide good approximation to the quarkonium, under which the Schr\"odinger-type equation can be employed to study the spectroscopy of the bound state. Recently, the lattice QCD studies released quantification of the in-medium spectrum--mass shift and thermal widths of Bottomonium ($b\bar b$) up to 3S and 2P states in QGP\cite{Larsen:2019zqv}, which was found can not be reproduced by the one-loop HTL motivated functional form of the heavy quark in-medium potential, $V_R(T, r)$ and $V_I(T, r)$. Note that the mass shift may give influence to quarkonium production in HICs~\cite{Chen:2012gg}.

In Ref.~\cite{Shi:2021qri} the authors devised a model-independent DNN-based method to reconstruct the temperature and inter-quark distance dependent in-medium heavy quark potential based upon the lattce QCD results mentioned above for Bottomonium. Since the universal approximation theorem, the DNN is introduced to parameterize the potential in an unbiased yet flexible enough fashion. The DNN represented heavy quark potential is coupled to the Schr\"odinger equation solving process to be converted into complex valued energy eigenvalues $E_n$, which are related to the bound state in-medium mass and thermal width through $\Re[E_n]=m_n-2m_b$ and $\Im[E_n]=-\Gamma_n$. By comparing to the lattice QCD “measurements”, the corresponding $\chi_2$ provide the loss function for optimizing the parameters of the potential-DNN, 
\begin{align}
\mathcal{L}=\frac{1}{2}\sum_{T, n}(\frac{m_{T, n} - m^{LQCD}_{T, n}}{\delta m^{LQCD}_{T, n}})^2 + (\frac{\Gamma_{T, n} - \Gamma^{LQCD}_{T, n}}{\delta\Gamma^{LQCD}_{T, n}})^2, \label{eq:hqloss}
\end{align}
with $T\in\{0, 151, 173, 199, 251, 334\}$ MeV and $n\in\{1S, 2S, 3S, 1P, 2P\}$ according to the lattice QCD evaluation conditions. Gradient descent with back propagation techniques can be applied for the DNN optimization here, of which the gradient is estimated efficiently from perturbative analysis on the Schr\"odinger equation with respect to perturbative change of the potential and just arrived in the Hellman-Feynman theorem. Furthermore, the uncertainty of the reconstructed potential is quantified by invoking Bayesian inference, thus evaluating the posterior distribution of the DNN parameters. With the outlined approach, Ref.~\cite{Shi:2021qri} achieved nice agreement with the lattice QCD results on masses and thermal widths of Bottomonium simultaneously, see left and middle panel of Fig.~\ref{fig.mass}. Meanwhile, the temperature and distance dependent heavy quark potential is also obtained as displayed in the right panel of Fig.~\ref{fig.mass}. Clearly the color screening effect emerged for the reconstruction with flatter structure appearing in $V_R(T, r)$ with increasing temperature at large distance, but the temperature dependence is found to be mild compared to perturbative analysis based results in the same temperature range considered. On the other hand, the imaginary part, $V_I(T, r)$, shows significant growth both with temperature and distance, and also shows greater magnitude than one-loop HTL motivated results. 

{\it{Deep Learning Quasi Particle Mass}} The equation of state of hadron resonance gas in the QCD phase diagram can be calculated using simple statistical formula with the following partition function,
\begin{align}
\ln Z(T) = \sum_i \ln Z_{hi}(T)
\end{align}
where $Z_{hi}(T)$ is the partition function for one of the several hundred hadrons in HRG, assuming that there is no interaction between different hadrons. The calculated EoS agrees with lattice QCD calculations. It is not possible to get the lattice QCD EoS for QGP using the 
same formula, as quarks and gluons interact with each other and form a many body quantum system. 
However, if one assumes that the quarks and gluons are non-interacting quasi particles whose masses depend on the local temperature, 
it is able to reproduce Lattice QCD EoS using the following simple statistical formula,
\begin{align}
\ln Z(T)= & \ln Z_g(T)+\sum_i \ln Z_{q_i}(T) \nonumber \\
\ln Z_g(T)= & -\frac{d_g V}{2 \pi^2} \int_0^{\infty} p^2 d p \nonumber\\
& \ln \left[1-\exp \left(-\frac{1}{T} \sqrt{p^2+m_g^2(T)}\right)\right] \nonumber\\
\ln Z_{q_i}(T)= & +\frac{d_{q_i} V}{2 \pi^2} \int_0^{\infty} p^2 d p \nonumber\\
& \ln \left[1+\exp \left(-\frac{1}{T} \sqrt{p^2+m_{q_i}^2(T)}\right)\right]
\label{eqs:partition}
\end{align}
where $Z_g$ is the partition function of quasi gluons, $Z_{q_i}$ represent the partition function of quasi quarks, $d_g$ and $d_{q_i}$ are the spin and color degeneracy for gluons and quarks respectively, $p$ is the magnitude of momentum, $T$ is the local temperature. Gluons, up, down and strange quarks are taken into account in this calculation. It is assumed that the temperature quasi particle masses $m_{u/d}(T)$ are the same for up and down quarks, but different for gluons $m_g(T)$ and strange quarks $m_s(T)$. As a result, there will be 3 variational functions whose forms are unknown and to be determined by mapping the following resulted EoS to Lattice QCD EoS,
\begin{align}
P(T)  = T\left(\frac{\partial \ln Z(T)}{\partial V}\right)_T \nn
\epsilon(T)  = \frac{T^2}{V} \left(\frac{\partial \ln Z(T)}{\partial T}\right)_V.
\label{eqs:stat_eos}
\end{align}

Several deep residual neural networks are constructed to represent the variational functions $m_{u/d}(T), m_s(T)$ and $m_g(T)$.
The mass functions of these quasi partons are used in Eq.~\ref{eqs:partition} to compute the partition function.
The resulted partition function is further used in Eq.~\ref{eqs:stat_eos} to compute the pressure and energy density as a function of temperature.
Notice that in this procedure there are both numerical integration and differentiation.
The integration is implemented using Gauss quadrature in the language of tensorflow, while the differentiation is given by auto-differentiation.
The loss function is designed as,
\begin{align}
{\rm loss} = |s_{\rm dnn} - s_{\rm lattice}|^2 + |\Delta_{\rm dnn} - \Delta_{\rm lattice}|^2 + L_{\rm constrain}
\label{eqs:loss_quasi}
\end{align}
where $s = {(\epsilon + P) / T}$ is the entropy density, $\Delta = {(\epsilon - 3 P) / T^4}$ is the trace anomaly.
The $L_{\rm constrain}$ contains physical constraints at high temperature region whose theoretical function form is given by HTL calculations.
The learned quasi partons reproduce Lattice QCD EoS. Using these mass functions, the authors calculated $\eta/s(T)$ and found that its minimum located around $1.25 T_c$ \cite{Li:2022ozl}.

\section{Hard Probe}
\label{sec:hard}

Energetic partons lose energy as they pass through the hot quark gluon plasma.
This process is quantified by the jet transport coefficient $\hat{q}$,
which is defined as the transverse momentum broadening squared per unit length \cite{Baier:1996kr,Baier:1996sk,Gyulassy:1993hr,Guo:2000nz,Wiedemann:2000za}.
The temperature dependent jet transport coefficient for heavy quarks is extracted using Bayesian analysis,
using the D-meson $v_2$ and $R_{AA}$ data from different experiments \cite{Xu:2017obm}.
Bayesian inference extracted the jet energy loss distributions, showing that the observed jet quenching
is dominated by only a few out-of-cone scatterings \cite{He:2018gks}.
The JETSCAPE collaboration extracted $\hat{q}$ with a multi-stage jet evolution approach model \cite{Soltz:2019aea}.
These studies typically use parametrized forms for the unknown $\hat{q}(T)$ function.
An information field is proposed to provide non-parametric functions for global Bayesian inference in order to
to avoid long-range correlations and human biases \cite{Xie:2022ght,Xie:2022fak}.

Deep learning has been widely used in high energy particle physics to analyse the substructures of jets and to 
to classify jets using the momentum of final state hadrons in jets \cite{Feickert:2021ajf,Du:SSPMA-2022}. 
In heavy ion collisions, deep learning is used not only to classify quark and gluon jets, but also to study the jet energy loss, the medium response and the initial jet production positions \cite{Du:2020pmp,Du:2021pqa,Yang:2022yfr}. 

Constraining the initial jet production positions will allow more detailed and differential studies of jet quenching. 
For example, one task in HIC is to search for Mach cones in QGP produced by the supersonic parton jet. The difficulty is that the jets are produced at different locations in the initial state and travel in different directions in the QGP. As a result, the shape of the Mach cone depends on the path length and is distorted by the local radial flow and temperature gradient. Predicting jet production positions using deep learning will help to select jet events whose Mach cones have a similar shape, thus enhancing the signal of the Mach cones in the final state hadron distribution. 

In these studies, the training data are usually generated by jet transport models \cite{He:2015pra,JETSCAPE:2017eso}, e.g., in the linear Boltzmann transport model (LBT), the jet parton loses energy through elastic scattering with thermal partons in QGP and inelastic gluon radiation. This process is described by a linearised Boltzmann equation, shown below,
\begin{align}
p_a &\cdot\partial f_a=\int \prod_{i=b,c,d}\frac{d^3p_i}{2E_i(2\pi)^3}\frac{\gamma_b}{2}(f_c f_d-f_a f_b)\left |\mathcal{M}_{ab\to cd} \right |^2 \nn
\\ 
&\times S_2(\hat{s},\hat{t},\hat{u})(2\pi)^4\delta^4(p_a+p_b-p_c-p_d)+\rm inelastic. \label{eq:LBT}
\end{align}
where $f_{a/c}$ are the distribution functions of the jet partons before and after scattering in the forward process, $f_{b/d}=1 / \left[ e^\frac{p\cdot u}{T} \pm 1\right]$ are the Fermi-Dirac and Bose-Einstein distributions for thermal quarks and gluons, respectively, in QGP. On the right hand side, $f_c f_d$ corresponds to the gain term and $-f_a f_b$ to the loss term during elastic scattering, whose amplitude squares $|\mathcal{M}_{ab\to cd} |^2$ from leading order perturbative QCD calculations. The $\gamma_b$ is the colour and spin degeneracy of the thermal parton $b$ and the term $\hat{S}_2 = \theta(\hat{s}>2 \mu_D^2)\theta(-\hat{s} + \mu_D^2 \le \hat{t} \le - \mu_D^2)$ is used to regularise the collinear divergence. The inelastic part comes from the gluon radiation described by higher-twist calculations \cite{}.

The lost energy is deposited in QGP as source terms of the relativistic hydrodynamic equations,
\begin{align}
\nabla_{\mu} T^{\mu \nu} = J^{\nu} \label{eq:CLVisc}
\end{align}
where $T^{\mu\nu}$ is the local energy momentum tensor of QGP and $J^{\nu}$ is the source term. In practice, if the energy deposited on the recoiled thermal parton exceeds $2$ GeV, it will be taken out and put into the LBT. This leaves a negative jet source in QGP. If the deposited energy is less than $2$ GeV, this corresponds to a positive jet source. Recoiled partons in the LBT do not interact with each other, which explains why the LBT solves a linearised Boltzmann equation. Recently, the LBT has been extended to the QLBT, which treats quarks and gluons as quasi-partons to constrain various transport parameters \cite{Liu:2021dpm}.

The initial jet production positions are sampled from the distribution of hard scattering, which is proportional to the distribution of the number of binary collisions. The initial entropy density distribution is provided by the Trento Monte Carlo model, from which the initial $T^{\mu\nu}$ can be calculated. Simultaneously solving Eq.~\ref{eq:LBT} and Eq.~\ref{eq:CLVisc} provides both the jet energy loss and the medium response in each simulation. Typically $10\sim 100$ thousand jet events are required to predict the initial jet production positions. Of course, the more training data the better, if there are sufficient computing resources.

One might ask whether there is a type of deep neural network that is better suited to studying jet energy loss and predicting jet production positions. In practice, convolutional neural networks (CNNs), point cloud neural networks, and graph neural networks have all been used in different projects. Typically, the performance of different neural network architectures is tested on these candidates and the one that works best for the specific task is selected. The simplest yet most powerful CNN should be the first to be tried in jet shape and jet energy loss studies. To capture the full information in jets, a point cloud network and a message passing neural network can be used.

\section{Observables in HIC}
\label{sec:obs}
{\it{PCA for flow analysis}}
In relativistic heavy-ion collisions, the collective flow provides important information about the properties of the QGP and its initial state fluctuations~\cite{Teaney:2009qa,Romatschke:2009im,Heinz:2013th,Gale:2013da,Song:2013gia,Song:2017wtw}. The flow observables are generally defined by a Fourier decomposition of the produced particle distribution in momentum space, such as:
\begin{equation}\label{eq:1}
\frac{{\rm d} N}{{\rm d} \varphi} =\frac{1}{2\pi}  \sum_{-\infty}^{\infty}\vec{V}_n e^{-in\varphi}=\frac{1}{2\pi} (1+ 2 \sum_{n=1}^{\infty} v_{n} e^{-in(\varphi-\Psi_n)})
\end{equation}
where  $\vec{V}_n =v_ne^{in\Psi_n}$ is the flow-vector of order n,  $v_{n}$ is  flow harmonics of order n and $\Psi_{n}$ is the corresponding event plane angle. The flow coefficients  can also be obtained from the two-particle correlations associated with a Fourier decomposition:
\begin{equation}\label{eq:1}
\lr{\frac{dN_{\mathrm{pairs}}}{d{p}_1 d{p}_2}} \propto 1+2\sum_{n=1}^{\infty} V_{n\Delta}(\pTa,\pTb) \cos (n \Dphi)
\end{equation}
where $V_{n\Delta}(\pTa,\pTb)$ is a symmetric covariance matrix and
$\Delta \phi=\phi^a-\phi^b$ is the
relative azimuthal angle between two emitted particles. Under the assumption of flow factorization, $V_{n\Delta}(\pTa,\pTb)$
is related to the flow harmonics $v_{n}(\pT)$ by: $V_{n\Delta}(\pTa,\pTb)\approx v_n(\pTa)v_n(\pTb)$ ~\cite{Gardim:2012im} (For other flow methods or flow measurements, see~\cite{Voloshin:2008dg,Snellings:2011sz,Jia:2014jca,Heinz:2013th,Song:2017wtw}).

\begin{figure}[t]
	\centering
		\includegraphics[width=1.0\linewidth,height=5cm]{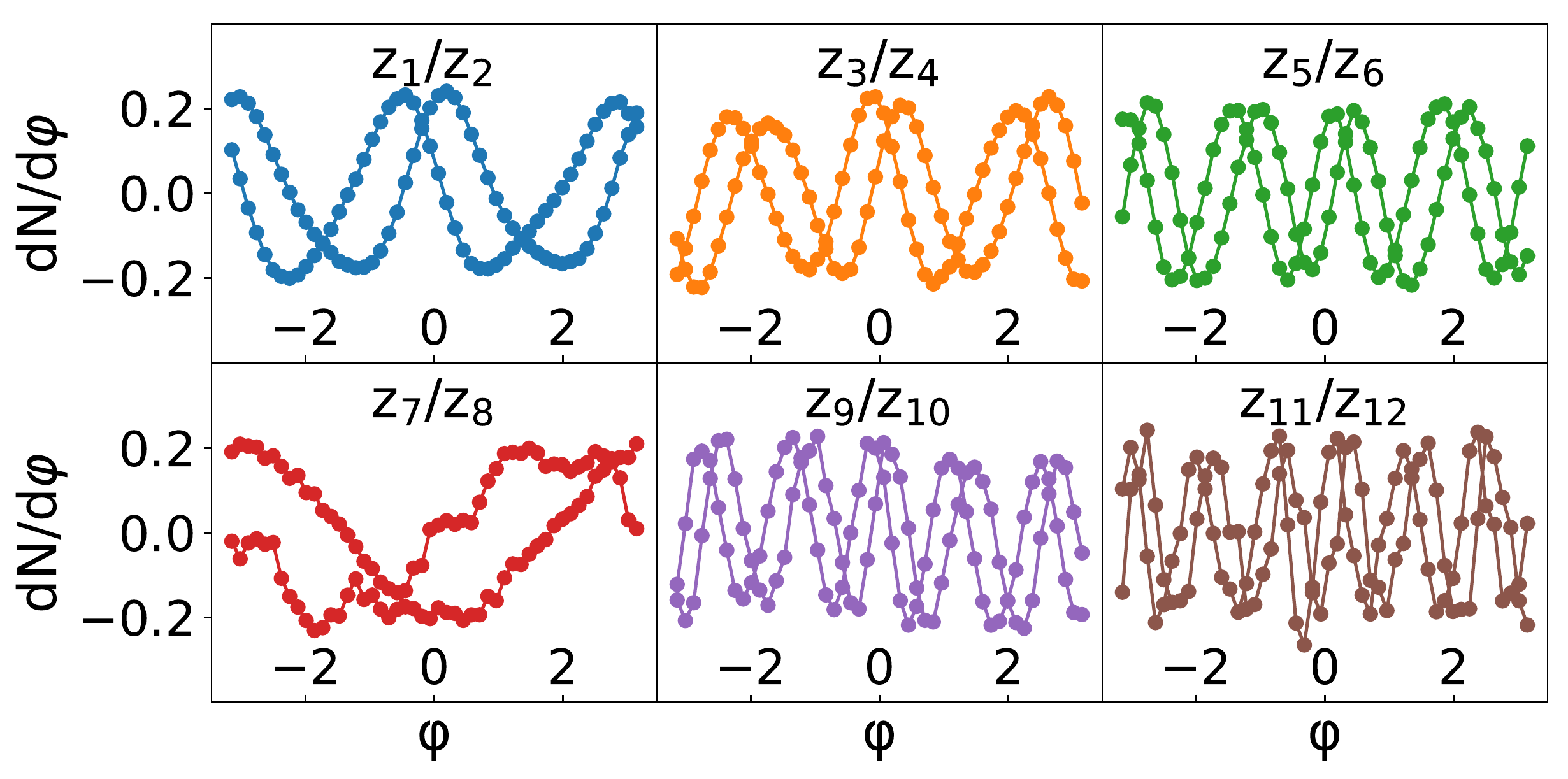}
\caption{The PCA eigenvectors ${z}_j$ for the final state matrix of particle distributions,  generated from {VISH2+1} hydrodynamics in 2.76 A TeV Pb+Pb collisions at 10\%-20\% centrality~\cite{Liu:2019jxg}.}
	\label{fig:eigenmodes1}
\end{figure}

\begin{figure}[t]
	\centering
		\includegraphics[width=0.8\linewidth,height=5cm]{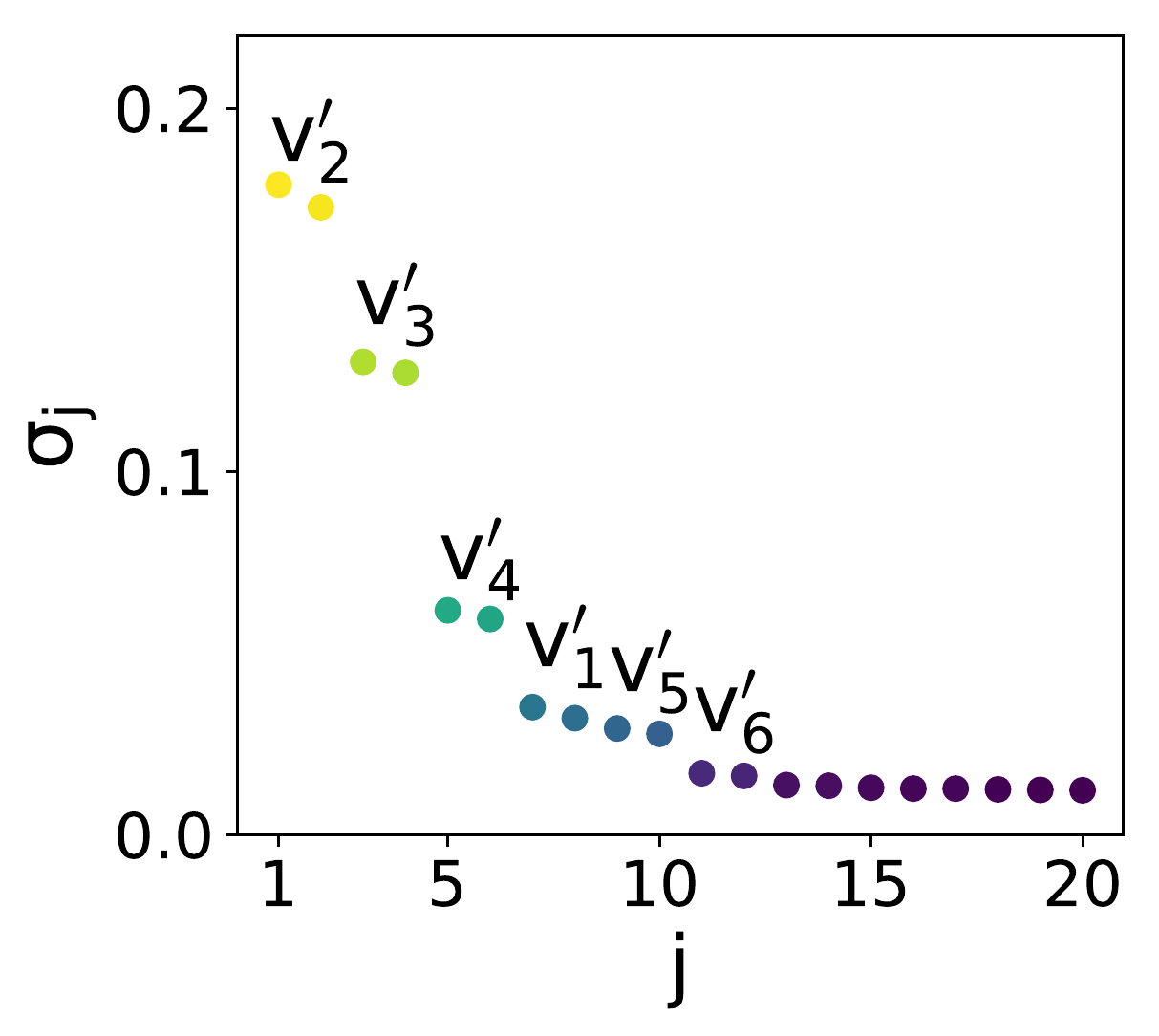}
\caption{The singular values of PCA for the final state matrix of particle distributions in Pb+Pb collisions at 10-20\% centrality~\cite{Liu:2019jxg}.}
	\label{fig:eigenmodes2}
\end{figure}

\begin{figure*}[t]
	\includegraphics[width=0.95\linewidth]{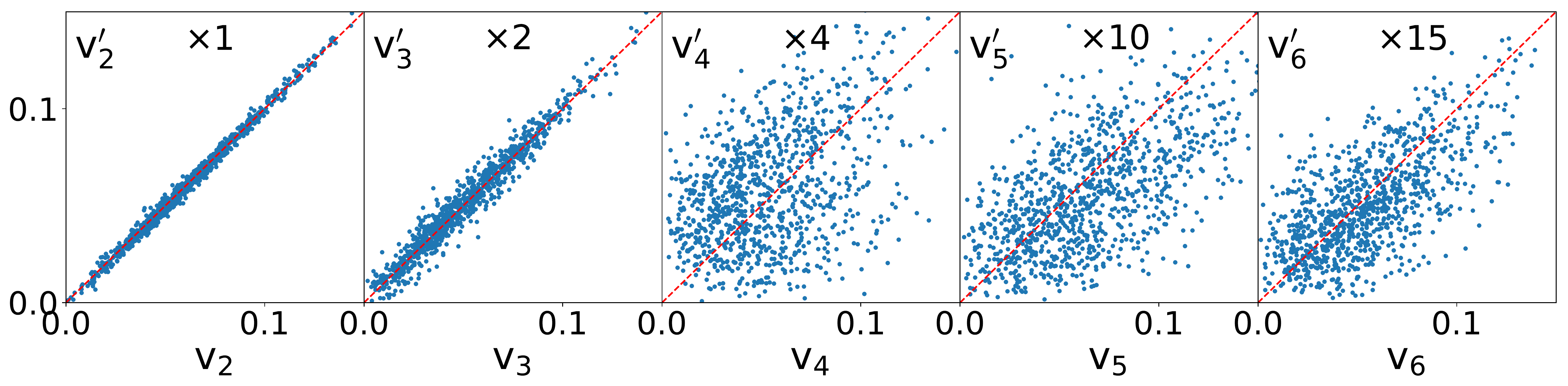} 
	\caption{A comparison between the  event-by-event flow harmonics $v_n'$ from PCA and $v_n$ from the Fourier expansion in Pb+Pb collisions at 10-20\% centrality~\cite{Liu:2019jxg}.}
	\label{fig:vv_compare}
\end{figure*}

Recently, one of the machine learning techniques, called the Principal Component Analysis (PCA) based on the Singular Value Decomposition (SVD), has been implemented to study the collective flow in relativistic heavy-ion collisions. For the 2-particle correlations with the Fourier expansion~\cite{Bhalerao:2014mua,Mazeliauskas:2015vea,Mazeliauskas:2015efa,Bozek:2017thv}, the event-by-event flow fluctuations have been investigated by PCA, which reveals the substructures of the flow fluctuations~\cite{Bhalerao:2014mua,Mazeliauskas:2015vea,Mazeliauskas:2015efa}. In more detail, with PCA, $V_{n\Delta}(\pTa,\pTb)$ can be expressed as~\cite{Mazeliauskas:2015vea}:
\begin{eqnarray}\label{eq:7}
&V_{n\Delta}(\pTa,\pTb)=\sum_{\alpha}v_n^{(\alpha)}(\pTa)v_n^{(\alpha)}(\pTb)\;,\\\label{eq:8}
\mathrm{with} \ &\int d\pT w^2(\pT) v_n^{(\alpha)}(\pT)v_n^{(\beta)}(\pT)=\lambda_\alpha\delta_{\alpha\beta}
\end{eqnarray}
where $v_n^{(\alpha)}(\pT)$ are the eigenvectors of the two-particle covariance matrix, $w(\pT)$ is the weight for the particle. $\alpha=1$ denotes the leading modes, $\alpha=2$ denotes the subleading modes and $\alpha=3$ denotes the subsubleading modes, and so on. It was found that the leading modes correspond to the traditional flow harmonics and the sub-leading modes lead to the breakdown of  the flow factorization. Using hydrodynamic simulations, Ref~\cite{Mazeliauskas:2015vea,Mazeliauskas:2015efa} demonstrated a linear relationship $V_n^{(\alpha)}\propto {\mathcal{E}}_{n}^{(\alpha)}$ for the leading, subleading and subsubleading modes. In Ref.~\cite{Bozek:2017thv}, PCA was implemented to study the mode coupling between flow harmonics, which found some hidden mode-mixing patterns that had not been discovered before. Recently, the CMS collaboration extracted the subleading flow modes for Pb+Pb and p+Pb collisions at the LHC, which showed a qualitative agreement between experimental measurements and the theoretical calculations~\cite{CMS:2017mzx}. Using {AMPT} and {HIJING} simulations, Ref.~\cite{Liu:2020ely}
showed that the PCA modes depend on the choice of the $\pT$ range and the particle weight $w$. In addition, the leading modes are affected by the non-flow effects, and the mixing between the non-flow and leading flow modes  leads to
fake subleading modes. Therefore, it is important to carefully handle with the non-flow effects and the choice of weight and phase space when implementing PCA to extract the subleading flow modes in both experimental and theoretical sides.

These above PCA studies of collective flow~\cite{Bhalerao:2014mua,Mazeliauskas:2015vea,
Mazeliauskas:2015efa,Bozek:2017thv,CMS:2017mzx,Liu:2020ely} are all based on the correlation data obtained with a Fourier expansion. Recently, PCA has been applied directly to the single particle distributions $dN/d\varphi$ without any prior treatment with a Fourier transform, which aimed to explore
whether PCA could directly discover flow without the guidance from human-beings~\cite{Liu:2019jxg}. More specifically, with a PCA
matrix multiplication, the $i_{th}$ row of a particle distribution matrix with N events, generated from {VISH2+1} hydrodynamics, can be expressed as:
\begin{eqnarray}
dN/d\varphi^{(i)}&=&\sum_{j=1}^m {x}_j^{(i)}{\sigma}_j {z}_j
=\sum_{j=1}^m \tilde{v}_j^{(i)} {z}_j  \nonumber  \\
&\approx & \sum_{j=1}^{{k}} \tilde{v}_j^{(i)} {z}_j \ \ \ (i)=1,... ,N
\label{pca}
\end{eqnarray}
Here, $(i)=1,2,... , N$  is the index of the event. $j$ is the index for the azimuthal angle where the total azimuthal angle $[-\pi,\pi]$ is divided into $m$ bins in order to count the number of particles in each bin. After the Singular Value Decomposition (SVD), $dN/d\varphi^{(i)}$ can be expressed by a linear combination of the eigenvectors $z_j$ with the corresponding coefficient $\tilde{v}_j^{(i)}$ (where $j=1,2,... ,m$), and $\sigma_i$ is the diagonal elements (singular values) of the particle distribution matrix, which is arranged in a descending order. In the spirit of PCA, in the last step,  a cut is made at the indices ${k}$ to focus  only on the most important components.

Fig.~\ref{fig:eigenmodes1} and Fig.~\ref{fig:eigenmodes2} show the first 12  eigenvectors ${z}_j $  and the first 20 singular values $\sigma_j$ of the PCA  in descending order for the final state matrix constructed from 2000 $dN/d\varphi$ distributions with the azimuthal angle $[-\pi,\pi]$ equally divided into 50 bins. Such $dN/d\varphi$ distributions are generated from the VISH2+1 hydrodynamics with event-by-event fluctuating  TRENTo initial conditions  for 2.76 A TeV Pb+Pb collisions at 10\%-20\% centrality. Fig.~\ref{fig:eigenmodes1} shows that the PCA eigenvectors are similar to the traditional Fourier bases. For example, the $1{st}$ and $2{nd}$ eigenvectors are close to $\mathrm{sin}(2\varphi)$ and $\mathrm{cos}(2\varphi)$ and the
$3{rd}$ and $4{th}$ eigenvectors are close to $\mathrm{sin}(3\varphi)$ and $\mathrm{cos}(3\varphi)$. The corresponding singular values in Fig.~\ref{fig:eigenmodes2} are also arranged in pairs,
which correspond to the real and imaginary parts of the anisotropic flow. It was found that, for $n\leq6$,
the values of these PCA flow harmonics are very close to those of the traditional event averaged flow harmonics obtained from the Fourier expansion, but not exactly the same. Fig.~\ref{fig:vv_compare} compares the event-by-event flow harmonics obtained from PCA and from the traditional Fourier expansion.  It shows that the elliptic flow with $n=2$ and the triangular flow with $n=3$ from both methods are in good agreement. However, for higher flow harmonics with $n \geq 4$, the PCA and Fourier expansion ones differ significantly from each other due to the mode mixing effects. With these PCA flow harmonics $v'_n$, Ref.~\cite{Liu:2019jxg} also calculates the symmetric cumulants  $SC{'(m,n)}$. Except for $SC{'(2,3)}$, these PCA symmetric cumulants largely decrease compared with the traditional Fourier ones, due to the largely increased linearity between the PCA flow harmonics and the initial eccentricities. These results indicate that PCA could define  the collective flow on its own basis. Compared with  the traditional ones obtained from the Fourier decomposition, the PCA method reduces the mode coupling effects between different flow harmonics~\cite{Liu:2019jxg}.  \\

{\it{CME detection}}
In the presence of magnetic field, chiral magnetic effect (CME) phenomenon can occur when the system shows chiral imbalance thus the number of left-handed and right-handed particles differs. Basically, a current of electric charge (known as chiral magnetic current) can be induced to flow along the direction of the magnetic field. It's been proposed to use Chiral magnetic effect to reveal the vacuum structure of QCD. In heavy ion collisions, a strong magnetic field can be created from the motion of the colliding ions, also it's predicted that in the formed hot and dense QGP the topological fluctuations of gluon fields may cause chiral imbalance for quarks, accordingly the CME may take place which can manifest as a separation of electric charge along the magnetic field direction. However, several challenges hinder the detection of CME in heavy ion collisions, among which the chief difficulty is to disentangle the CME signal from other possible sources of charge separation e.g., elliptic flow, the global polarization and other background noises though multiple observables are proposed.

\begin{figure}[htbp!]
    \includegraphics[width=8.0 cm]{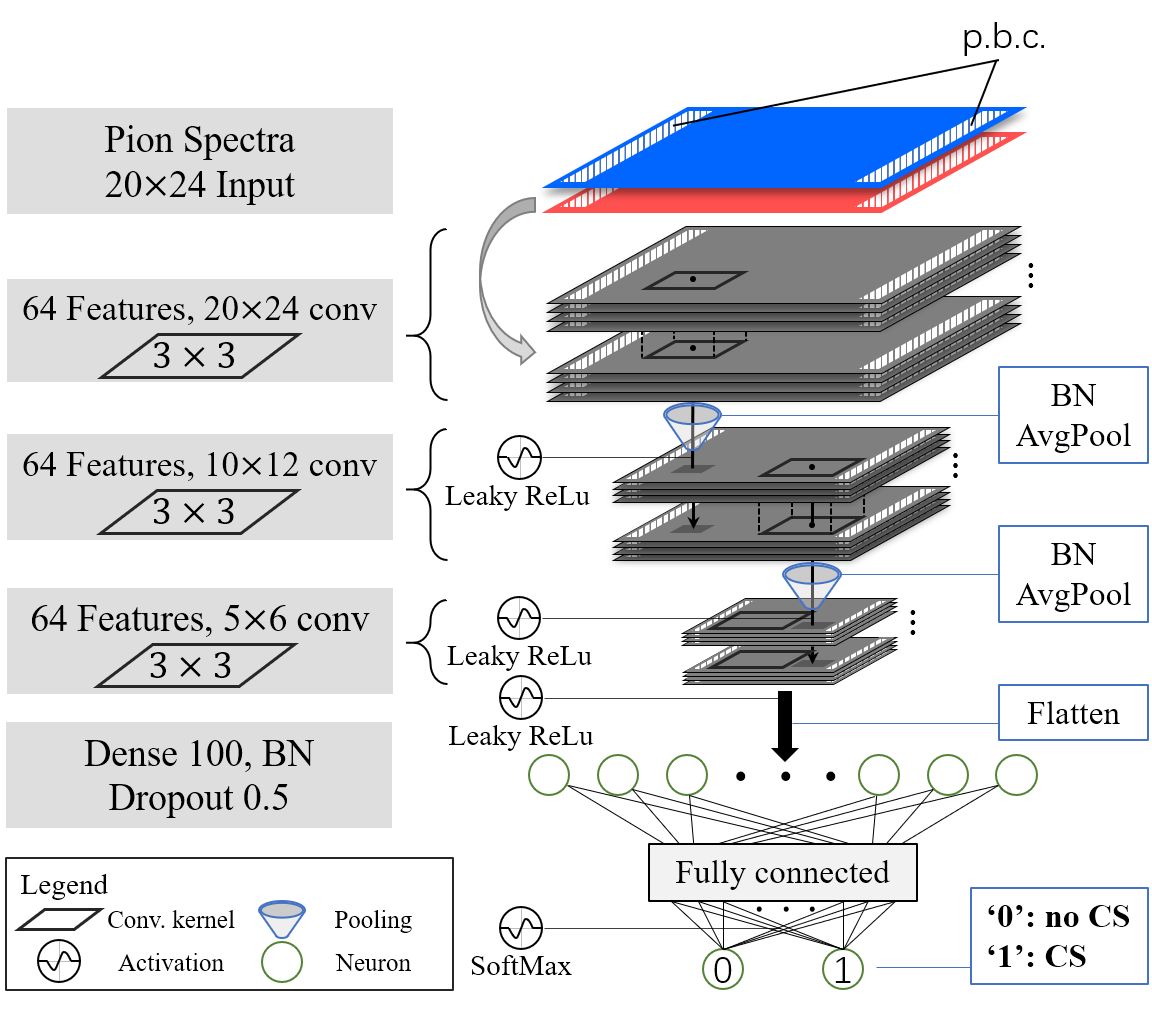}
    \setlength{\belowcaptionskip}{-0.cm}
    \caption{Taken from ~\cite{Zhao:2021yjo}. The convolutional neural network architecture with $\pi^+$ and $\pi^-$ spectra $\rho^{\pm}(p_T, \phi)$ as input.
    }
    \label{fig:CME_CNN}
\end{figure}

Despite the challenges, there's a long-term and continuing interest for the search for CME in heavy ion collisions since its general importance to QCD. Recently in Ref.~\cite{Zhao:2021yjo} it's proposed to use deep learning to construct an end-to-end CME-meter, which can efficiently analyze the final-state hadronic spectrum as a whole in the sense of big data with deep convolutional neural network to disclose the fingerprints of CME. In performing supervised learning, the training set is prepared from the string melting a multiphase transport (AMPT) model with CME implemented under a global charge separation (CS) scheme. Basically, the CME events are generative by switching the y-components of momenta of a fraction of downward moving light quark, and it's corresponding anti-quarks with upward moving direction. The fraction defines the CS fraction, $f$ which separates the events to “no CS” (label as “0”) class for those with $f=0\%$ and “CS” class (label as “1”) for those with $f>0\%$. Each event is represented as two-dimensional transverse momentum and azimuthal angle spectra of charged pions in the final state, $\rho_{\pi}(p_T,\phi)$.
Then deep CNN is trained to perform binary classification on the labeled events with the spectra to be the input. See Fig.~\ref{fig:CME_CNN} about the architecture of the devised deep CNN for CME-meter construction.

As seen from Fig.~\ref{fig:CME_CNN} the output of the network has two nodes, with each of them being naturally interpreted as the probability resulting from the network decision in recognizing any given input spectrum as CME ($P_1$) or non-CME ($P_0=1-P_1$) events. 
The training set contains multiple collision beam energies and centralities for diversity consideration.  The pion spectra is obtained by averaging over 100 events with the same collision condition to reduce the fluctuations, which also reduces the backgrounds and thus should be considered as prerequisite for realistic application in experiment. For the training, different levels of CS fraction is used, and it's found that the classfiction validation accuracy for using smaller CS fraction training events is less than the model trained with higher CS fraction events. This shows that larger CS fraction can be identified easier, which is natural. Despite the different induced discernibility, the trained deep CNNs all showed robust performance against varying collision centrality and energy. One can conclude that at least under the AMPT modelling level the CS signals can survive into the final state of the collision dynamics at different collision conditions, which can be recognized by the deep CNN-based CME-meter.
\begin{table}[htbp!]
\caption{The results of the (0\%+10\%) model on the isobaric collision systems (Ru+Ru and Zr+Zr at 200 GeV).}
\label{tab:isobar}
\centering
\begin{tabular}{ccccccc}
\hline
Centrality\quad & 0-10\% \quad & 10-20\% \quad & 20-30\% \quad & 30-40\% \quad & 40-50\% \quad & 50-60\% \\ \hline
$R_{\rm iso}$  \quad & 9.95\% \quad & 12.99\% \quad & 8.13\% \quad & 13.84\% \quad & 19.67\% \quad & 10.47\% \\ \hline
\end{tabular}
\end{table}

Note that the network is trained just on Au+Au collision systems, while the extrapolation to other collision system is showed to work successfully. Specifically, the obtained CME-meter is applied to isobaric collisions of $_{40}^{96}$Zr+$_{40}^{96}$Zr and $_{44}^{96}$Ru+$_{44}^{96}$Ru, which is proposed for the search of CME especially. Since the Ru contains more protons to induce a larger magnetic field than Zr, it's expected that there would be a larger CS signal in Ru+Ru collisions. To reveal this difference and also the distinguishable difference of the CME-meter on the two isobaric collision systems from $P_1^\text{Ru}>P_1^\text{Zr}$, the $R_{iso}$ is evaluated which well verify the developed CNN based CME-meter,
\begin{equation}
    R_\text{iso}=2\times\frac{\langle\text{logit}(P_1^\text{Ru})\rangle-\langle\text{logit}(P_1^\text{Zr})\rangle}{\langle\text{logit}(P_1^\text{Ru})\rangle+\langle\text{logit}(P_1^\text{Zr})\rangle},
\end{equation}
where the function $\text{logit}(x)=\text{log}[{x}/{(1-x)}]$ is used to restore the derivative at saturation region of the activation in NN last layer, SoftMax. From Tab.~\ref{tab:isobar} on $R_{iso}$ it's seen that the trained CME-meter gets well verified beyond the training collision system, which indicate its robust capture of general CME signal in the collisions.

The CME-meter is also validated on a different model simulation, anomalous-viscous fluid dynamics (AVFD).  $P_1$ shows consistent positive correlation along with increasing $N_5/S$ which controls the CME strength, while the contamination from local charge conservation (LCC) up to $30\%$ didn't augment the performance of the CME-meter on the testing events from AVFD. To reveal the underlying account for the trained CME-meter, in Ref.~\cite{Zhao:2021yjo} the comparison between network output $P_1$ and $\gamma$-correlator is also investigated, where the $\gamma$-correlator as conventional CME probe can measure the event-by-event two-particle azimuthal correlation of charged hadrons. It's shown that on averaged events, both the CME signal and the background from $\delta\gamma$ (difference between correlations within same charged particles and correlation within opposite charged particles) get suppressed. Being differently, the CME-meter output $P_1$ works well in classifying CS and no-CS classes on the averaged events.

The direct implementation of this trained CME-meter into real experiments would require firstly reconstructing the reaction plane of each collision event to form the averaged events as input for the meter. In general the reaction plane reconstruction can be reached by measuring correlations of final state particles which inevitably contains finite resolution and background effects. It's shown that even on restricted event plane reconstruction, the trained CME-meter still can recognize the CS signals. To render the deployment of the trained CME-meter on single event measurements, Ref.~\cite{Zhao:2021yjo} also proposed a hypothesis test perspective.

\begin{figure}[htbp!]
    \centering
    \includegraphics[width=7 cm]{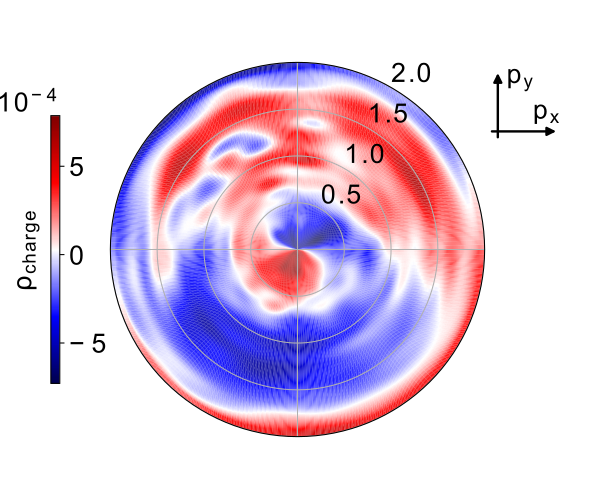}
    \caption{\textit{DeepDream} map for the (0\%+10\%) model~\cite{Zhao:2021yjo}.}
    \label{fig:DeepDream}
\end{figure}
Another way to interpret the trained deep learning algorithm is the DeepDream method, which was used in Ref.~\cite{Zhao:2021yjo} to reconstruct the network most responding input pion spectrum, manifesting the “CME pattern” that the CNN-based CME-meter essentially captured for its further CME signal recognition. The key idea is to perform variational tuning on the input pion spectrum with the trained and frozen network to maximize its output (i.e., pushing $P_1\to 1$), driven by the gradient $\delta P_1(\rho_{\pi}(p_T,\phi))/\delta\rho_{\pi}(\p_T,\phi)$. The resultant “CME pattern” from the trained network is displayed in Fig.~\ref{fig:DeepDream}, from which the charge conservation and a clear dipole structure appears, both being CME related features.

\section{Summary and Outlook}
\label{sec:sum}
{\it{Summary}}
As a modern computational paradigm, artificial intelligence (AI) especially the machine- and deep-learning techniques are nowadays opening up a wealth of applications and new possibilities to the forefront of scientific studies. Due to its special ability in recognizing patterns and structures hidden in complex data, these learning algorithms based strategies make it feasible for physics exploration with big data or a smart computation mindset. In the context of high energy nuclear physics revolving around heavy ion collision programs to understand nuclear matter properties under different conditions, different research topics are benefiting as well from the incorporation of these techniques.

We present the recent progresses in the area of heavy ion collisions in this mini-review article, including initial states physics inference, QCD matter transport and bulk properties study, thermal medium modifications for parton or hadrons as well as physical observables recognition in HICs.

For methodology, we first overview different loss functions $l$ used in supervised learning, un-supervised learning, semi-supervised learning, self-supervised learning and active learning. During training, the negative gradients $-\frac{\partial l}{\partial \theta}$ are used to optimize the network in SGD-like algorithms. Auto differentiation is employed to compute the derivatives of the loss with regard to model parameters $\theta$, efficiently. As auto-diff has analytical precision for the variational function represented by the neural network, it has been widely used in physics informed neural network to solve ODEs and PDEs. We then introduced the widely used neural network architectures, such as MLP, CNN, RNN and Point Cloud Network. Afterward, we explained the generative models such as Auto Encoder, GAN, flow model and diffusion models in details.
These models are widely used in Lattice QCD to generate field configurations. 

For the initial condition, machine learning has been widely used to determine the centrality classes (impact parameter) using the final state hadrons in momentum space, to extract the initial nuclear structures such as the nuclear deformation, the $\alpha$ clustering as well as the neutron skin. In general, it seems  easier to extract the nuclear  deformation than the $\alpha$ clustering and neutron skin in current literature.

For bulk matter, Bayesian parameter estimation has been successfully used to determine the temperature dependent shear and bulk viscosity of QGP.
Unsupervised auto encoder is used to reconstruct the charged multiplicity distributions, which helps to determine the source temperature and the temperature of nuclear liquid gas phase transition.
Deep CNN, point cloud network and event averaging techniques are employed to classify the crossover and first order phase transition regions in QCD phase diagram, using data generated with relativistic hydrodynamic models and hadronic transport models.
Active learning is used to mapping out thermodynamically unstable regions near the critical endpoint in the QCD phase diagram.   For hydrodynamic evolution, a well-designed network, called sUnet, could capture the non-linear mapping between initial and final profiles with sufficient precision, which is also much faster than the traditional hydrodynamic simulations.

For QGP in-medium effects, we first reported some of the recent machine learning based spectral function reconstruction, which is a notorious ill-posed inverse problem. Both supervised and unsupervised methods have been discussed for the inference of spectral out of Euclidean correlator measurements from Monte Carlo simulation (e.g., lattice study). Then the in-medium heavy quark interaction inference based upon in-medium heavy quarkonium spectroscopy is introduced, where a novel DNN representation integrated inside the forward problem-solving pipeline with automatic differentiation approach is proposed. This strategy is also used then for in-medium quasi particle effective model construction from the lattice QCD EoS.

For hard probes, Bayesian analysis is widely used to extract the temperature dependent jet (or heavy quark) transport coefficient $\hat{q}(T)$ 
and the jet energy loss distributions.
Recently, deep learning assisted jet tomography is developed to locate the initial jet production positions.
This is important for the study of jet substructures and the medium response.
Using this technique, it was observed that the signal of jet induced Mach cones is amplified by selecting jet events.

For the observables, the Principal Component Analysis (PCA) has been implemented to study the collective flow in relativistic heavy ion collisions. It revealed the substructures of the flow fluctuations, which can potentially be implemented to extract the subleading modes of flow with the efforts from both experimental and theoretical sides.
After applied directly to the single particle distributions, PCA could directly discover flow with a basis similar to the Fourier expansion ones, which also greatly reduces the mode coupling between different flow harmonics.

{\it{Outlook}}
In spite of many impressive progresses, the interplay between high energy nuclear physics and machine learning is still inducing hectic evolution. Yet, many questions and challenges exist and deserve further exploration. Besides the previously mentioned applications of ML into heavy ion collisions, several foreseeable topics could be explored with ML paradigm as well, e.g., for critical end point searching, in eRHIC and EIC regime~\cite{Lee:2022kdn}, spin polarization study, the upcoming FAIR program, nuclear structure inference, HIAF experiment in China, etc.
About the future prospects on this field of applying machine learning techniques for heavy ion collision physics study, since it's fast evolving and still under developing, we here throw out several questions those we think deserve future efforts, to get the ball rolling to advance our field:

\begin{itemize}

\item{Can ML define us more efficient "observables” to pin down the desired physics?}

\item{Could the algorithm dig out new physical knowledge to advance our understanding of nuclear matter from the data? }

\item{How to make those ML algorithms to be confronted with realistic experiment? if it's possible to do on-line analysis? How to access experimental raw data to test neural network pretrained with model simulations?}

\item{If it's possible to speed up heavy ion collision dynamical simulations for confronting high statistic measurements or in Bayesian inference?}

\item{How to combine Bayesian inference strategy with machine learning to advance our field and better connect experiment to theory?}

\item{How we fully take symmetries into the analysis using ML? e.g., Lorentz Group Equivariant Autoencoders \cite{Hao:2022zns}, other symmetries? How would dimensionality analysis (constraints) be incorporated properly inside the ML methods consistently?}
\end{itemize}

It's also worthy to think about how we expand more potentially useful approaches from other fields' development, e.g., from particle physics, condensed matter physics, astrophysics, or even other scientific disciplines, and how the community to better organize with joint efforts, like taking deeper look into the direction and maximize the potential of these novel computational techniques to really advance the field of HENP.

\begin{acknowledgments}
This work was supported in part by the National Natural Science Foundation of China under contract Nos. 11890714 and 12147101 (Ma), 12075098 (Pang),  12247107 and 12075007 (Song), and the Germany BMBF under the ErUM-Data project (Zhou), and Guangdong Major Project of Basic and Applied Basic Research No. 2020B0301030008 (Ma).
\end{acknowledgments}

\bibliography{reviews}
\end{document}